\documentclass[global,numbook]{svjour}
\input{psfig}

\newcommand {\boxfigure}[1]%
   {\framebox[\textwidth]{%
    \parbox {0.99\textwidth}
                {{#1}\vspace {0cm}\hfill}}}

\newcommand {\boxfigureone}[1]%
   {\framebox[\textwidth]{%
    \parbox {0.90\textwidth}
                {{#1}\vspace {0cm}\hfill}}}

\newcommand{\gefig}[3]
	{\begin{figure}[htb] 
        \begin{center}%
	\mbox{}
	{\psfig{figure=#1,width=0.60\textwidth}}
	\mbox{}
	\end{center}
	\caption{{#2}\label{#3}} 
	\end{figure}}

\newcommand{\wgefig}[4]
        {\begin{figure}[htb] 
        \begin{center}%
        \mbox{}
        {\psfig{figure=#1,width=#4}}
        \mbox{}
        \end{center}
        \caption{{#2}\label{#3}} 
        \end{figure}}

\newcommand{\cgefig}[2]
        {\begin{figure}[htb] 
        \begin{center}%
        \mbox{}
        {\psfig{figure=#1,width=0.60\textwidth}}
        \mbox{}
        \end{center}
        \label{#2} 
        \end{figure}}

\newcommand{\eat}[1]{}

\newcommand{\imp}{{\ :\!\!-\ }}
\newcommand{\mc}{{mc\_carthy\_91}}

\begin{document}
\newtheorem{observation}[theorem]{Observation}
\title{A General Framework for Automatic \\ Termination Analysis
of Logic Programs\thanks{This research has
been partially supported by  grants from the Israel Science Foundation}}

\author{Nachum Dershowitz\inst{1} 
\and Naomi Lindenstrauss\inst{2} 
\and Yehoshua Sagiv\inst{2}
\and Alexander Serebrenik\inst{3}}
\institute{School of Computer Science,
Tel-Aviv University, Tel-Aviv 69978, Israel.
\email nachum@cs.tau.ac.il 
\and Institute for Computer 
Science, The Hebrew University, Jerusalem 91904, Israel.
\email \{naomil,sagiv\}@cs.huji.ac.il
\and
 Department of Computer Science, K.U. Leuven,
                Celestijnenlaan 200A, B-3001 Heverlee, Belgium.
\email Alexander.Serebrenik@cs.kuleuven.ac.be}

\titlerunning{General Framework for Automatic Termination Analysis}
\maketitle
\begin{abstract}

This paper describes a general framework for automatic termination
analysis of logic programs, where we understand by ``termination'' the finiteness
of the LD-tree constructed for the program and a given query.
A general property of mappings from a certain subset of the branches of
an infinite LD-tree into a finite set  is proved. From this result several
termination theorems are derived, by using different finite sets.
The first two are formulated for the
predicate dependency and atom dependency graphs. Then a general result
for the case of the query-mapping pairs relevant to a program
is proved (cf. \cite{Sagiv,Lindenstrauss:Sagiv}). The
correctness of the {\em TermiLog}
system described in \cite{Lindenstrauss:Sagiv:Serebrenik} follows from it. 
In this system it is not possible to prove 
termination for programs involving arithmetic predicates, since the usual 
order for the integers is not well-founded. A new method, which can be easily
incorporated in {\em TermiLog} or similar systems, 
is presented, which  makes it possible
to prove  termination for programs involving arithmetic predicates. 
 It is based on combining a finite abstraction of the integers with the 
technique of the query-mapping pairs, and is essentially capable of dividing 
a termination proof into several cases, such that a simple termination function
suffices for each case. Finally several possible extensions are outlined.


\end{abstract}

\section{Introduction}
The results of applying the ideas of abstract interpretation to logic
programs (cf. \cite{Cousot:Cousot}) 
seem to be especially beautiful and useful, because we are
dealing in this case with a very simple language which
 has only one basic construct---the
clause. Termination of programs is known to be undecidable, but again
things are simpler for logic programs, because the only possible cause
for their non-termination is infinite recursion, so it is possible to
prove termination automatically for a large class of programs. 
For a formal proof 
of the undecidability of the termination of general logic programs see~\cite{Apt:Handbook}.

The kind of termination we address is the termination of the computation of
all answers to a goal, given a program, when we use Prolog's computation
rule (cf. \cite{Lloyd:Book}). This is equivalent to finiteness of the  LD-tree
constructed for the program and query (the LD-tree is the SLD-tree constructed
with Prolog's computation rule---cf. \cite{Apt:Book}). Even if one is
interested only in a single answer, it is important to know that computation
of all answers terminates, since the solved query may be backtracked into
(cf. \cite{OKeefe}).

One of the difficulties when dealing with the LD-derivation of a goal, given
a logic program, is that infinitely many non-variant atoms may appear as
subgoals. The basic idea is to abstract this possibly infinite structure
to a finite one.
We do this by mapping partial branches of the LD-tree to the elements of a
finite set of abstractions $\cal A$. By using the basic lemma of the paper
and  choosing different possibilities for $\cal A$, we get different results
about termination. The first two results are formulated for the predicate
dependency and atom dependency graphs.

Then we get, by using the query-mapping pairs of \cite{Sagiv,Lindenstrauss:Sagiv}, first
a termination condition that cannot be checked effectively and then a
condition that can. The latter forms the core of the {\em TermiLog}
system (cf. \cite{Lindenstrauss:Sagiv:Serebrenik}), 
a quite powerful system we have developed for checking termination
of logic programs. 
 
Then a new method, which can be easily incorporated in the {\em TermiLog} 
or similar systems, is presented for showing termination
for logic programs with arithmetic predicates. Showing termination in this case
is not easy, since the usual order for the integers is not well-founded. 
The method consists of the following steps:
First, a finite abstract domain for representing the range of integers
is deduced automatically. Based on this abstraction, abstract 
interpretation is applied to the program.  The result is a finite 
number of atoms 
abstracting answers to queries, which are used to extend the 
technique of query-mapping pairs. For each query-mapping 
pair that is potentially non-terminating, 
a bounded (integer-valued) termination function is guessed.
If traversing the pair decreases the value of the termination function,
then termination is established.
Usually simple functions  suffice for each query-mapping
pair, and that gives our approach an edge over the classical approach of 
using
a single termination function for all loops, which must inevitably
be more complicated 
and harder to guess automatically.
It is worth noting that the termination of McCarthy's
91 function can be shown automatically using our method.

Finally generalizations of the
  algorithms presented  are pointed out, which make it possible to deal
 successfully with even more cases.

\section{Preliminaries}
Consider the LD-tree determined by a program and goal.
\begin{definition}Let $\leftarrow r_1,\ldots , r_n$ and
$\leftarrow s_1, \ldots , s_m$ be two nodes on the same branch of the LD-tree,
with the first node being above the second. We say  $\leftarrow s_{1},
\ldots ,s_{m}$  is a {\em direct offspring} of  $\leftarrow r_{1},
\ldots ,r_{n}$ if $s_{1}$
is, up to a substitution, one of the body atoms of the clause with which 
$\leftarrow r_1 ,\ldots , r_n $ was resolved. 
 We define the {\em offspring} relation as
 the irreflexive transitive closure of the direct offspring relation. We  call
a path between two
  nodes in the tree such that one is the offspring of the other
a {\em call branch}.
\end{definition}

Take for example the {\tt add-mult} program given in Figure~\ref{add-mult-ex}
and the goal {\tt mult(s(s(0)),s(0),Z)}.
\begin{figure}[htb]
\begin{verbatim}
(i)     add(0,0,0).
(ii)    add(s(X),Y,s(Z)) :- add(X,Y,Z).
(iii)   add(X,s(Y),s(Z)) :- add(X,Y,Z).
(iv)    mult(0,X,0).
(v)     mult(s(X),Y,Z)) :- mult(X,Y,Z1), add(Z1,Y,Z).
\end{verbatim}
\caption{add-mult example}
\label{add-mult-ex}
\end{figure}

The LD-tree  is given in Figure~\ref{add-mult-LD}.
\begin{figure}[htb]
\setlength{\unitlength}{1mm}
\begin{picture}(100,80)(0,0)
\put(0,70){(1) $\leftarrow mult(s(s(0)),s(0),Z)$}
\put(0,65){(2) $\leftarrow mult_{(1)}(s(0),s(0),Z1), add_{(1)}(Z1,s(0),Z)$}
\put(0,60){(3) $\leftarrow mult_{(2)}(0,s(0),Z2), add_{(2)}(Z2,s(0),Z1), add_{(1)}(Z1,s(0),Z)$}
\put(15,55){$\{Z2 \mapsto 0\}$}
\put(0,50){(4) $\leftarrow add_{(2)}(0,s(0),Z1), add_{(1)}(Z1,s(0),Z)$}
\put(15,45){$\{Z1 \mapsto s(Z3)\}$}
\put(0,40){(5) $\leftarrow add_{(4)}(0,0,Z3), add_{(1)}(s(Z3),s(0),Z)$}
\put(15,35){$\{Z3 \mapsto 0\}$}
\put(0,30){(6) $\leftarrow add_{(1)}(s(0),s(0),Z)$}
\put(15,25){$\{Z \mapsto s(Z4)\}$}
\put(65,25){$\{Z \mapsto s(Z5)\}$}
\put(0,20){(7) $\leftarrow add_{(6)}(0,s(0),Z4)$}
\put(50,20){(8) $\leftarrow add_{(6)}(s(0),0,Z5)$}
\put(15,15){$\{Z4 \mapsto s(Z6)\}$}
\put(65,15){$\{Z5 \mapsto s(Z7)\}$}
\put(0,10){(9) $\leftarrow add_{(7)}(0,0,Z6)$}
\put(50,10){(10) $\leftarrow add_{(8)}(0,0,Z7)$}
\put(15,5){$\{Z6 \mapsto 0\}$}
\put(65,5){$\{Z7 \mapsto 0\}$}
\put(0,0){(11) $\leftarrow $}
\put(50,0){(12) $\leftarrow $}
\end{picture}
 \caption{LD-tree}
\label{add-mult-LD}
\end{figure}
In this case node (2) and node (6) are, for instance, direct offspring of
node (1), because the first atoms in their respective goals 
come from the body of  clause (v),
 with which the goal of node (1) was resolved.
Note that we add to the predicate of each atom in the LD-tree a subscript
that denotes who its `parent' is, i.e., 
the node in the LD-tree that caused this atom
to be called as the result of resolution.
A graphical representation of the direct offspring relation is given in
Figure~\ref{offspring}.

\begin{figure}
\setlength{\unitlength}{1mm}
\begin{picture}(110,40)(25,0)

\put(80,30){(1)}
\put(60,20){(2)}
\put(100,20){(6)}
\put(50,10){(3)}
\put(70,10){(4)}
\put(90,10){(7)}
\put(110,10){(8)}
\put(70,0){(5)}
\put(90,0){(9)}
\put(110,0){(10)}

\put(78,29){\line(-2,-1){12}}
\put(98,22){\line(-2,1){12}}
\put(59,19){\line(-1,-1){5}}
\put(99,19){\line(-1,-1){5}}
\put(69,12){\line(-1,1){5}}
\put(109,12){\line(-1,1){5}}
\put(72,8){\line(0,-1){5}}
\put(92,8){\line(0,-1){5}}
\put(112,8){\line(0,-1){5}}
\end{picture}
\caption{The offspring relation}
\label{offspring}
\end{figure}

The following theorem holds:
\begin{theorem}
\label{offspring-theorem}
  If there is an infinite branch in the LD-tree
corresponding to a program and query then there is an infinite sequence
of nodes $N_{1},  N_{2}, \ldots $ such that for each $i$, $N_{i+1}$ is 
an offspring of $N_{i}$.
\end{theorem}
\begin{proof}
Straightforward.
\end{proof}

The main idea of the paper is to find useful finite sets of
abstractions of call branches
 and to formulate termination results in terms of them.
An effort has been made to make the presentation as simple and
self-contained as possible.

\section{The  basic lemma}
Given an LD-tree we define a shadow of it as a mapping from its set of call
branches to a finite set of abstractions.
\begin{definition}[Shadow]
Let an LD-tree for a query and program and a finite set $\cal A$ be given.
A {\em shadow} of the LD-tree into $\cal A$ is a mapping $\alpha$ that assigns to each call
branch of the tree an element of $\cal A$.
\end{definition}
Then the following basic lemma holds
\begin{lemma}[Basic Lemma]
\label{Basic:Lemma}
 Suppose the LD-tree for a program and a query has an infinite branch.
 Let $\alpha$ be a shadow mapping from  the call branches of the tree
into a finite set $\cal A$. Then there is a sequence of  nodes 
$M_{1},M_{2},\ldots $ and an element $A\in {\cal A}$, such that for each
$i$, $M_{i+1}$ is an offspring of $M_{i}$,
and for each $j,k$ the call branch from $M_{j}$ to $M_{k}$ is mapped by $\alpha$ to $A$.
\end{lemma}
\begin{proof}
By Theorem~\ref{offspring-theorem}, there is an infinite sequence of nodes 
$N_{1},  N_{2}, \ldots $, such that for each $i$, $N_{i+1}$ is
an offspring of $N_{i}$. To each call branch from $N_i$ to an $N_j$ the 
mapping $\alpha$ assigns one of the elements of the finite set ${\cal A}$.
By Ramsey's theorem~\cite{Graham} we get that there is a subsequence
$N_{k_1},  N_{k_2}, \ldots $, such that for each $i,j$ the mapping $\alpha$
assigns to the branch from $N_{k_i}$ to $N_{k_j}$ the same element.
\end{proof}

There is some structure in the set of call branches.
If we have two call branches, one going from $N_{1}$ to $N_{2}$ and one
going from $N_{3}$ to $N_{4}$, we can if $N_{2}=N_{3}$ define their
composition, which is the branch from $N_{1}$ to $N_{4}$. This operation
is associative. In accordance with the nomenclature in algebra we can
call a set $S$ with a partial associative operation $* : S\times S
\rightarrow S$ a semi-groupoid. We may want the finite set $\cal A$ to be
a semi-groupoid too and the mapping $\alpha$ to be a homomorphism. This
brings us to the definition of a structured shadow.
\begin{definition}[Structured Shadow] Let an LD-tree for a query and
program and a finite semi-groupoid $\cal A$ be given. A structured shadow of
the LD-tree into $\cal A$ is a mapping $\alpha$ that assigns to each
call branch of the tree an element of $\cal A$ so that for any two
call branches  $B_{1}$ and $B_{2}$ that can be composed
we have that $\alpha (B_{1} )*\alpha (B_{2} )$
 is defined and
\[\alpha (B_{1}*B_{2}) = \alpha (B_{1} )*\alpha (B_{2} )\]
\end{definition}

When defining a structured shadow it is enough to give the value of
$\alpha$ for call branches between nodes and their direct offspring.
This is the reason for the name.

The element $A$ whose existence is proved in the basic lemma is, in the
case of a structured shadow, an element that can be composed with itself.
We call such an element a {\em circular element}. Moreover, it is {\em idempotent}.

\section{Two simple applications of the basic lemma}
In the following sections we'll give applications of the basic lemma. In each
case we'll give the set of abstractions $\cal A$ which will always be
finite and the mapping $\alpha$ from
call branches to elements of $\cal A$.
In the first two applications we use the absence of circular elements in
$\cal A$ to derive termination.
   
\subsection{The Predicate Dependency Graph}
Take as $\cal A$  elements of the form $(p\rightarrow q)$ where $p$ and $q$
are predicate symbols of the program. Define composition as
\[(p\rightarrow q)*(q \rightarrow r)=(p\rightarrow r)\]
If a call branch goes from a node $\leftarrow p(X_{1},\ldots ,X_{n}),\ldots$
to another node $\leftarrow q(Y_{1},\ldots ,Y_{m}),\ldots$ we'll define the value of
$\alpha$ on it as $(p\rightarrow q)$.
It is not difficult to see that $\alpha$ is a structured shadow.

The {\em predicate dependency graph} of a program is a graph whose vertices
are the predicate symbols of the program and such that for each clause
$A\imp B_{1},\ldots ,B_{n}$ it has an arc from the predicate of $A$
to the predicate of $B_{i}$, for $i=1,\ldots ,n$ (cf. \cite{Plumer:Book}).
If $N_{1},N_{2}$ are nodes in the LD-tree such that one is the
direct offspring of
the other then the value of $\alpha$ for the call branch between them can
be seen as an arc in the predicate dependency graph of the program.

  From the
basic lemma we get that if there is non-termination then there must be
a circular element in the image under $\alpha$ of all the call branches
of the LD-tree, that is, an element of the form $p \rightarrow p$. This means
that there is a non-trivial strongly connected component in the predicate
dependency graph (a trivial strongly connected component is one that
consists of a single vertex with no arc going from it to itself). Consequently,
the following well-known theorem follows from the basic lemma:
\begin{theorem}
 If there is no non-trivial strongly connected
component in the
predicate dependency graph of a program any query to it terminates.
\end{theorem}

It is easy to find examples of programs such that every query to them
terminates and yet their predicate dependency graph has non-trivial
strongly connected components.
 Take the  program

\begin{eqnarray*}
&& \tt at(jerusalem,mary).\\
&& \tt at(X,jane) \imp at(X,mary).
\end{eqnarray*}

\noindent
where the predicate dependency graph has the single vertex $at$ with an
arc $at\rightarrow at$.

\subsection{The Atom Dependency Graph}
Define
 two atoms to be equivalent if they are variants of each other.
For an atom $At$ denote by $[At]$ its equivalence class under variance.
Take as $\cal A$ elements of the form $[p(T_{1},\ldots ,T_{n})]
\rightarrow [q(S_{1},\ldots ,S_{m})]$ where $p(T_{1},\ldots ,T_{n})$
is an atom that appears in the head of a
clause in the program and $q(S_{1},\ldots ,S_{m})$ is
 an atom  that appear in the body of a clause. Composition
is defined for pairs

\noindent$[p(T_{1},\ldots ,T_{n})]\rightarrow [q(S_{1},\ldots ,S_{m})]$ and
$[q(R_{1},\ldots ,R_{m})]\rightarrow [r(W_{1},\ldots ,W_{k})]$

such that representatives that are named apart of
$[q(S_{1},\ldots ,S_{m})]$ and $[q(R_{1},\ldots R_{m})]$ can be unified. In that
case the result of the composition is
$[p(T_{1},\ldots ,T_{n})]\rightarrow  [r(W_{1},\ldots ,W_{k})]$.

Now suppose a node $\leftarrow p(T_{1},\ldots ,T_{n}),\ldots$ has as
direct offspring
  a node $\leftarrow q(S_{1},\ldots ,S_{m}),\ldots$ and suppose the clause
  used for resolution with the first node was $A\imp B_{1},\ldots , B_{l}$
  and that the atom $q(S_{1},\ldots , S_{m})$ originates in $B_{j}$. Then
  we will take $\alpha$ to map the call branch between the two nodes to $[A]\rightarrow [B_{j}]$.
For call branches between  nodes that are not direct offspring of
each other we define the value of $\alpha$
by composition.

We can define the {\em atom dependency graph} of a program as follows.
 Consider
 a graph whose vertices are equivalence classes of atoms that appear in the
 program.  If there is a rule $A\imp B_1, B_2, ..., B_n$
 then we put in the graph arcs  $[A]\rightarrow [B_{i}]\; (i=1,...,n)$.
We call these arcs ``arcs of the first kind''.
 Now if there are arcs of the first kind
$ [A_{1}]\rightarrow [A_{2}]$  and $ [B_{1}]\rightarrow [B_{2}]$
and named apart variants of  $A_{2}$  and  $B_{1}$  can be
 unified, we also add an arc  $[A_{2}]\rightarrow [B_{1}]$. Such an arc we call ``an arc of the second kind''. The graph we get we call the atom dependency
 graph (note the similarity to the U-graph of \cite{wang}). For the example at the end of the previous subsection the atom
 dependency graph consists of the two vertices $[at(X,jane)],[at(X,mary)]$
 and an arc from the first to the second.

From the basic lemma we get that if there is an infinite branch in the
LD-tree there must be a circular element in the image under $\alpha$ of
the call branches of the LD-tree.

So we get the following conclusion of the basic lemma:
\begin{theorem}
 If there is no non-trivial strongly connected component in the
atom dependency graph of a program any query to it terminates.
\end{theorem}

Again it is not difficult to find programs such that every query to them
terminates and yet their atom dependency graph has non-trivial
strongly connected components.
Take the following program

\begin{eqnarray*}
&&\tt p(X,f(Z)) \imp q(X,f(Z)).\\
&&\tt q(g(Y),W) \imp r(g(Y),W).\\
&&\tt r(X,X) \imp p(X,X).
\end{eqnarray*}

\noindent for which every SLD-tree is finite, but for which there is a strongly connected component consisting of 6 nodes
 in its atom dependency graph---if
we denote by $a,b,c,d,e,f$ respectively the atom dependency graph nodes 
\[ [p(X,f(Y))],\; [q(X,f(Y))],\; [q(g(X),Y)],\; [r(g(X),Y)],\; [r(X,X)],\; [p(X,X)]\]
then there are arcs of the first kind from $a$ to $b$, from $c$ to $d$ and from
$e$ to $f$ and arcs of the second kind from $b$ to $c$, from $d$ to $e$ and
from $f$ to $a$, so the nodes $a,b,c,d,e,f$ form a strongly connected component.

\section{The Abstraction to Query-Mapping Pairs}
We will now consider a more complex abstraction and take as $\cal A$
the set of  query-mapping pairs determined by the program.
In this case termination will follow not from the absence of circular elements
in the image of the shadow mapping, but from the absence of circular elements of a certain kind.

We start with a formal definition of query-mapping pairs. The meaning of the pairs
will be clarified later.
\begin{definition}[Mixed Graph]
A {\em mixed graph} is a graph with both edges and arcs.
(We use the usual terminology---edges are undirected, while arcs are directed.)
\end{definition}
 A  query-mapping pair  consists of two parts, both of which
are mixed graphs, however a different notation is used for each.
\begin{definition}[Query-Mapping Pair]
A {\em query-mapping pair} $(\pi , \mu )$ consists of two parts:
\begin{itemize}
\item The query $\pi$, that is a mixed graph
whose nodes correspond to argument positions of some predicate in the
program and are either black, denoted by {\bf b}, or white, denoted by
{\bf f}. An edge from the $i$'th to the $j$'th position will be denoted
by $eq(i,j)$. An arc from the $i$'th to the $j$'th position will be denoted
by $gt(i,j)$. As an example of a query for the {\tt add-mult} program
take {\sl mult(b,b,f) [gt(1,2),eq(2,3)]}.
\item The mapping $\mu$, that is a mixed graph whose nodes correspond to the
argument positions of the head of some rule (the {\em domain}) and the
argument positions of some body atom of that rule (the {\em range}). Again nodes can
be black or white. In this case we
depict the graph pictorially, as in Figure~\ref{mult-mapping}.
\wgefig{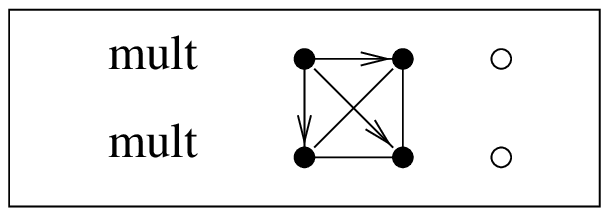}{Example of mapping}{mult-mapping}{0.40\textwidth}
\eat{
\vspace{2mm}
\begin{picture}(150,50)(2,0)
\put(50,20){mult}
\put(50,0){mult}
\put(90,20){\circle*{3}}
\put(110,20){\circle*{3}}
\put(130,20){\circle{3}}
\put(90,0){\circle*{3}}
\put(110,0){\circle*{3}}
\put(130,0){\circle{3}}
\put(91,20){\vector(1,0){18}}
\put(90,19){\vector(0,-1){18}}
\put(91,19){\vector(1,-1){17}}
\put(110,19){\line(0,-1){18}}
\put(109,19){\line(-1,-1){18}}
\put(91,0){\line(1,0){18}}
\end{picture}
\vspace{2mm}
} 
\end{itemize}
\end{definition}

For examples of query-mapping pairs see Figures~\ref{one-two}, \ref{six-eight}, \ref{six-ten}.

Clearly the number of query-mapping pairs that can be created by using the predicate symbols of a program is finite.

The means for proving termination is choosing a well-founded order on terms
and using it to show that the LD-tree constructed for the program and query
 cannot have infinite branches. Different
orders may be defined (cf. \cite{Dershowitz}). 
One of the ways an order can be given is by
defining a norm on terms. For example, one can 
use symbolic linear norms, which include as
special cases the term-size norm and the list-size norm. These symbolic
linear norms will be linear expressions, which we will be able to use
in the termination proof when they become integers. Essentially edges and
arcs will denote equality and inequality of norms and a node will become
black if its symbolic linear norm is an integer.

\begin{definition}[Symbolic Linear Norm]
 A {\em symbolic linear norm} for the terms created from an alphabet
consisting of function symbols and variables
is defined for non-variable terms by
\[ \| f(X_{1}, \ldots X_{n})\| = c+ \sum_{i=1}^{n} a_{i}\| X_{i}\|\]
where $c$ and $a_{1},\ldots a_{n}$ are non-negative integers that depend only
on $f/n$. This also defines the norm of constants if we consider them  as 
function symbols with $0$ arity. With each logical variable we associate
an integer variable to represent its norm (we use the same name for both, 
since the meaning of the variable is clear from the context). 
\end{definition}
\begin{definition}[Instantiated Enough]
 A term is {\em instantiated enough}
with respect to a symbolic linear norm if the expression giving its symbolic norm is
an integer.
\end{definition}
In this way of defining symbolic norms we follow \cite{van:Gelder}.
Some authors
define the norm of a variable to be $0$ and then use the norm only for terms
that are {\em rigid} with respect to it (cf. \cite{Decorte:DeSchreye:Fabris}). In our context it
is more convenient to use the symbolic norm. If the symbolic linear norm of a term
is an integer then we know that the term is rigid with respect to this particular
norm.

We get the term-size norm, which can be defined for a ground term as the number
of edges in its representation as a tree, or alternatively as the sum of the
arities of its functors, by setting for every $f/n$
\[c=n\;\; , \;\;  a_{1}=\cdots a_{n}=1\]
So, for instance, the symbolic term-size norm of $f(g(X,X,Y),X)$ is $5+3X+Y$.
The symbolic term-size norm of a term is an integer exactly when the term is
 ground.

To get the list-size norm we set for the list functor
\[\|[H|T]\|=1+\|T\|\]
that is $c=1,a_{1}=0,a_{2}=1$, and for all other functors equate the norm of a
 term with them as head functor to $0$. 
In this case the norm is a positive integer exactly
for lists that have  finite positive length, irrespective of whether their elements
are fully instantiated or not.

This is perhaps the place to note that, since for the term-size norm all the
$a_{i}$'s are nonzero, a term is instantiated enough with respect to it
only if it is ground, while for other symbolic norms a term may be instantiated
enough without being ground.

Given the LD-tree of a program we define the shadow mapping $\alpha _{real}$
as follows.

With each call branch between  nodes that are offspring of each other we associate a query-mapping
pair in the following way:
\newpage
\noindent
 If
\begin{itemize}
\item the node nearest to the root among the branch nodes  
is $\leftarrow p_{1},\ldots ,p_{m}$,
\item  the node farthest from the root is $\leftarrow q_{1},
\ldots ,q_{n}$,
\item the substitution $\theta$ is the composition of the substitutions associated
with the branch, 
\item $abs$ is the abstraction function which associates
with each atom the same atom with its arguments replaced by {\bf b} for 
arguments that are instantiated enough for the norm used
 and {\bf f} otherwise,
\end{itemize}
 then 
\begin{itemize}
\item the query of the pair is $abs(p_{1})$,
with the constraints that hold between the arguments of $p_{1}$,
\item the
domain of the mapping is $abs(p_{1}\theta)$, 
\item the range is $abs(q_{1})$, 
\item edges connect elements in the domain and range for which the corresponding
elements in the tree (i.e. the arguments of $p_{1}\theta$ and $q_{1}$)
 have the same norm,
\item arcs
connect elements in the domain and range for which the corresponding
elements in the tree are instantiated enough and for which
a norm inequality can be inferred.
\end{itemize}
 (The reader might be puzzled why we 
introduce arcs between elements for which a norm inequality can be inferred
{\em only} if the arguments are instantiated enough---the term-size of $s(X)$
will always be larger than that of $X$, whatever the substitution for $X$ will be.
However, to prove termination we use the well foundedness of the non-negative
integers, so will use the fact that there cannot be an infinite path of arcs,
since in our case they connect elements with integer norm.)

Take for example the {\tt add-mult} program given in Figure~\ref{add-mult-ex}
and the goal {\tt mult(s(s(0)),s(0),Z)} and
use the term-size norm (recall that for a ground term its term-size is the 
number of edges in its representation as a tree).

 To give a few examples of 
the query-mapping pairs we get for the LD-tree of this program and goal,
which is shown in Figure~\ref{add-mult-LD} (where we denote the constraints of
 a query by a list of elements of the form $eq(i,j)$ if the $i$'th and $j$'th
arguments have the same term-size, and $gt(i,j)$ if the term-size of the $i$'th
argument is greater than that of the $j$'th argument; and where there is in the 
mappings an edge between nodes with the same term-size and an arc from a node with
larger integer term-size to a node with smaller integer term-size):

For the call branch between node (1) and its direct offspring (2) we get the pair depicted in 
Figure~\ref{one-two}.
For the call branch between node (6) and its direct offspring (8) we get the pair presented in Figure~\ref{six-eight}.
For the call branch between node (6) and its offspring node (10) we get the pair in Figure~\ref{six-ten}. 
\wgefig{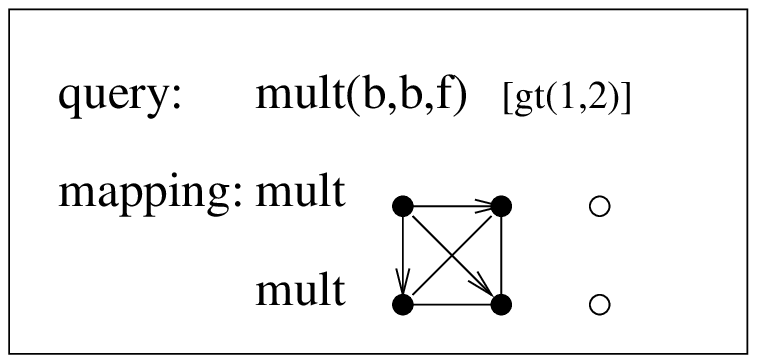}{Query-mapping pair for the branch from (1) to (2)}{one-two}{0.40\textwidth}
\wgefig{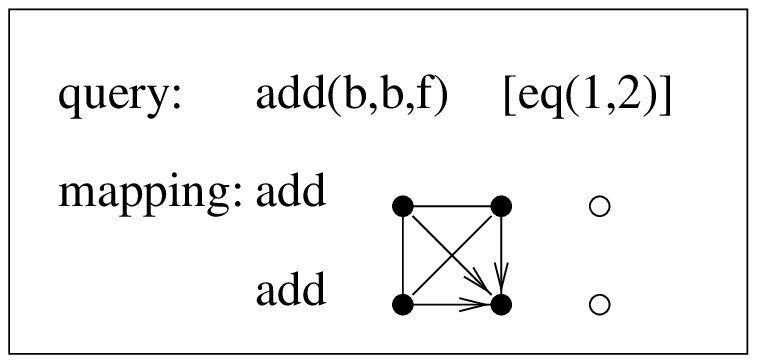}{Query-mapping pair for the branch from (6) to (8)}{six-eight}{0.40\textwidth}
\wgefig{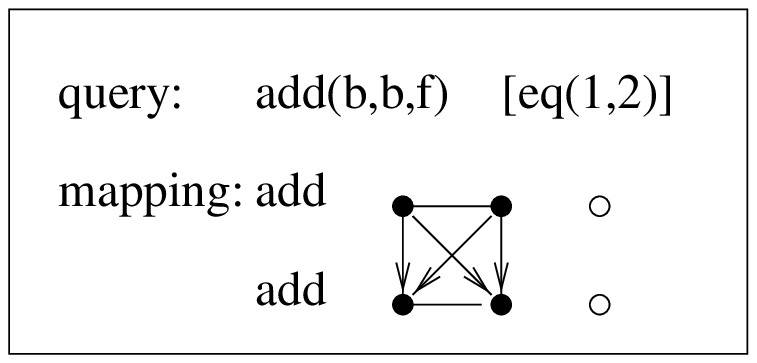}{Query-mapping pair for the branch from (6) to (10)}{six-ten}{0.40\textwidth}
\eat{ 
\vspace{2mm}
\fbox{
\begin{picture}(150,50)(2,0)
\put(0,40){query:}
\put(50,40){mult(b,b,f)}
\put(110,40){[l(2,1)]}
\put(0,20){mapping:}
\put(50,20){mult}
\put(50,0){mult}
\put(90,20){\circle*{3}}
\put(110,20){\circle*{3}}
\put(130,20){\circle{3}}
\put(90,0){\circle*{3}}
\put(110,0){\circle*{3}}
\put(130,0){\circle{3}}
\put(91,20){\vector(1,0){18}}
\put(90,19){\vector(0,-1){18}}
\put(91,19){\vector(1,-1){17}}
\put(110,19){\line(0,-1){18}}
\put(109,19){\line(-1,-1){18}}
\put(91,0){\line(1,0){18}}
\end{picture}
}
\vspace{2mm}
} 
\eat{ 
\vspace{2mm}
\fbox{
\begin{picture}(150,50)(2,0)
\put(0,40){query:}
\put(50,40){add(b,b,f)}
\put(110,40){[eq(1,2)]}
\put(0,20){mapping:}
\put(50,20){add}
\put(50,0){add}
\put(90,20){\circle*{3}}
\put(110,20){\circle*{3}}
\put(130,20){\circle{3}}
\put(90,0){\circle*{3}}
\put(110,0){\circle*{3}}
\put(130,0){\circle{3}}
\put(90,21){\line(1,0){18}}
\put(90,21){\line(0,-1){18}}
\put(91,19){\vector(1,-1){18}}
\put(110,19){\vector(0,-1){18}}
\put(109,19){\line(-1,-1){18}}
\put(91,0){\vector(1,0){18}}
\end{picture}
}
\vspace{2mm}
} 
\eat{ 
\vspace{2mm}
\fbox{
\begin{picture}(150,50)(2,0)
\put(0,40){query:}
\put(50,40){add(b,b,f)}
\put(110,40){[eq(1,2)]}
\put(0,20){mapping:}
\put(50,20){add}
\put(50,0){add}
\put(90,20){\circle*{3}}
\put(110,20){\circle*{3}}
\put(130,20){\circle{3}}
\put(90,0){\circle*{3}}
\put(110,0){\circle*{3}}
\put(130,0){\circle{3}}
\put(91,20){\line(1,0){18}}
\put(90,19){\vector(0,-1){18}}
\put(91,19){\vector(1,-1){18}}
\put(110,19){\vector(0,-1){18}}
\put(109,19){\vector(-1,-1){18}}
\put(91,0){\line(1,0){18}}
\end{picture}
}
\vspace{2mm}
} 

Note the following properties of query-mapping pairs in the image of $\alpha_{real}$.
A black node corresponds to an argument position which is instantiated enough
for the chosen symbolic norm to be an integer. A white node corresponds
 to an argument position that is potentially not instantiated enough. An edge connects
two nodes that have equal symbolic norms, and hence must be of the same
color. An arc goes from a  black node to another black node that has smaller
norm (recall that norms of black nodes  are non-negative integers).
The proof of termination uses the existence of such arcs and the 
well-foundedness of the non-negative integers with the usual order.

We define consistency of a mixed graph:
\begin{definition}[Consistency]
 A mixed graph is {\em consistent} if it has no positive cycle (i.e. a cycle 
that may contain both edges and arcs, but has at least one arc).
\end{definition}
Then it is clear that queries and mappings in the image must be
consistent.
A positive cycle may have only black nodes. This means that for
each argument $T$ represented by such a node we have that $\|T\|$ is an integer
and $\|T\|<\|T\|$, which is impossible. That is, a query-mapping pair that is not
consistent represents a branch that cannot really occur.

Note further that sets of edges and arcs for queries and mappings are closed under transitive closure
and that the domain of a mapping of a pair is subsumed by the query, where
subsumption is defined as follows.
\begin{definition}[Subsumption]
Given two mixed graphs with black and white nodes $G_{1}$ and $G_{2}$,
we say that $G_{1}$ is subsumed by $G_{2}$ if they have the same nodes up
to color,
every node that is black in $G_{2}$ is also black in $G_{1}$, and
every edge or arc between nodes in $G_{2}$ also appears
 for the respective nodes in  $G_{1}$.
\end{definition}

Among all query-mapping pairs we distinguish the `recursive' pairs, 
those for which the query is
identical to the range of the mapping (the  query-mapping pair given 
in Figure~\ref{six-ten} is of this kind). 

Before proceeding we need the following two definitions from \cite{Sagiv}.
We want to model recursive calls, so in the case of a query-mapping pair 
 $(\pi , \mu )$ such that $\pi$ is identical to the range of $\mu$, we create
what we call a {\em circular variant} by introducing special edges between
corresponding nodes in the domain and range. These special edges behave like
ordinary edges, except that they can only be traversed from range to domain.
What a circular edge between a range node and domain node models, is that the range
node can become unified with the domain node of another instance of the pair we
considered.
\begin{definition}[Circular Variant]
 If $(\pi , \mu )$ is a query-mapping pair, 
such that $\pi$ is identical to the range of $\mu$, then the {\em circular variant}
 of $(\pi , \mu )$ is  $(\pi , \mu ')$, where $\mu '$ is obtained from $\mu$ by 
connecting each pair of corresponding nodes in the domain and range with a {\em circular}
edge. 
\end{definition}
\begin{definition}[Forward Positive Cycle]
 A circular variant $(\pi , \mu )$ has a
{\em forward positive cycle} if $\mu$ has a positive cycle, such that when this cycle
is traversed, each circular edge is traversed from the range to the domain.
\end{definition}

From the basic lemma we get that if there is an infinite branch in the
tree, there must be an infinite sequence of nodes  $N_{1},N_{2},
\ldots$ such that for each $i$ the branch from $N_{i}$ to $N_{i+1}$ is
mapped into the same recursive pair. Suppose the circular variant of this
pair has a positive forward cycle. Start with a node on a forward positive cycle,
in the domain of the pair
corresponding to some call branch, say from $N_{i}$ to $N_{i+1}$. Now
traverse the cycle, but in the case of a circular edge go from the range of
this pair to the domain of the pair for the call branch from $N_{i+1}$ to
$N_{i+2}$. After a number of steps equal to the number of circular arcs on
the forward positive cycle we'll return to the same node in the pair as we
started from, only now corresponding to a lower call branch. From the 
existence of the forward positive cycle we can deduce a decrease
in norm. This means that we can find an infinite
sequence of arguments of atoms in the tree, such that their norms
form a descending sequence of non-negative integers, which is
impossible. So we get the following conclusion from the basic lemma

\begin{theorem} 
\label{LS-real-termination-criterion}
Let the LD-tree for a query and a program and a symbolic linear norm be given.
Define $\alpha _{real}$ as above. If all the circular
variants that can be created from the
query-mapping pairs in the image of $\alpha _{real}$
 have a forward positive cycle,
then the tree must be finite, i.e., there is termination for the query with
Prolog's computation rule.
\end{theorem}

Again it is easy to find an example of a program that terminates although it
does not satisfy the condition of the theorem. Take the program

\begin{eqnarray*}
&&\tt p(0).\\
&&\tt p(1) \imp p(0).
\end{eqnarray*}
with query pattern $p(f)$ and a norm that assigns the same value to $0$ and $1$.

Theorem~\ref{LS-real-termination-criterion}
does not give us an effecive way to determine
 for a particular program and query
if there is termination, because we cannot always construct the LD-tree that
may be infinite. The next section gives a way to approximate the `real' 
query-mapping pairs. The algorithm proposed either says that there is 
termination, or that it is not strong enough to decide.

\subsection{The Query-Mapping Pairs  Algorithm}
\label{qm-pairs-algo}
The following algorithm has been implemened in a system called {\em TermiLog}
(cf. \cite{Lindenstrauss:Sagiv,Lindenstrauss:Sagiv:Serebrenik}), 
which is quite powerful and has been able to analyze
correctly 82\% of the 120 benchmark programs it was
 tested  on, taken from the
literature on termination and other sources.
The basic idea of the algorithm  is to {\em approximate} the set of query-mapping pairs that
are associated with the LD-tree for a query and
program. We will show
that each `real' query-mapping pair arising from the LD-tree (i.e. the image
under $\alpha _{real}$ of a call branch) is subsumed by a query-mapping
pair in the approximation, so that a sufficient condition for the finiteness of the
LD-tree is that every circular variant in the approximation has a forward positive cycle.

We will define a structured shadow $\alpha _{app}$,
which is a {\em widening} (cf. \cite{Cousot:Cousot}) of $\alpha _{real}$,
by giving its value for call
branches between nodes that are direct offspring of each other (the
{\em generation} step) and  find its value for other call branches
by composition (the {\em composition} step).
It should be noted that in this case $\alpha _{app}$ associates with each call branch
a query-mapping pair that depends not only---as in the previous section---on the
nodes at the ends of the branch and the substitution associated with them, but
also on the location of the branch in the tree. Since we are approximating the
`real' pairs our conclusions are sound, but there may be constraints which
we have not inferred, so it may happen that the value of $\alpha _{app}$ for two call
branches which look identical but are in different parts of the tree will be
different.

The first step is constructing from each rule of the program a weighted rule graph, 
which extracts the information about argument norms that is in the rule.

\begin{definition}[Weighted Rule Graph] 
The {\em weighted rule graph} associated with a
rule has as nodes all the argument positions of the atoms in the rule; it has edges
connecting the nodes of arguments which have equal norm and has a potential 
weighted arc between any two nodes such that the difference between the norms of the respective
arguments, which is a linear expression, has non-negative coefficients and a
positive constant term, and this potential arc is labeled by the difference.
\end{definition}

 In our example, using the term-size norm, we get for the rule
\begin{eqnarray*} 
&& \tt mult(s(X),Y,Z)) \imp mult(X,Y,Z1), add(Z1,Y,Z).
\end{eqnarray*}
\noindent
the weighted rule graph that is shown in Figure~\ref{wrg}.
\wgefig{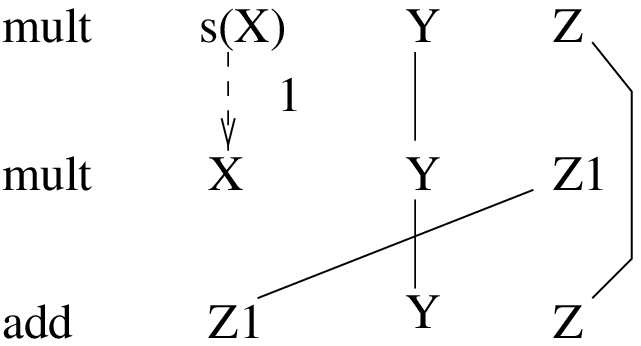}{Weighted rule graph}{wrg}{0.40\textwidth}
\eat{ 
\setlength{\unitlength}{1.1mm}
\begin{picture}(100,50)(5,-5)
\put(10,40){mult}
\put(40,40){s(X)}
\put(70,40){Y}
\put(100,40){Z}
\put(10,20){mult}
\put(40,20){X}
\put(70,20){Y}
\put(100,20){Z1}
\put(10,0){add}
\put(40,0){Z1}
\put(70,0){Y}
\put(100,0){Z}
\put(99,17){\line(-4,-1){53}}
\put(72,38){\line(0,-1){13}}
\put(72,18){\line(0,-1){13}}
\put(42,38){\line(0,-1){3}}
\put(42,33.5){\line(0,-1){3}}
\put(42,29){\vector(0,-1){4}}
\put(44,31){1}
\put(105,20){\oval(5,40)[r]}
\end{picture}
} 
\noindent The potential arc is shown by a dashed arc.
It should be explained what {\em potential} arcs are. In the termination proof 
we use the fact that the order induced by the norm on terms that are instantiated
enough is well-founded (recall that for such terms the norm is a non-negative
integer). Once we know that the nodes connected by a potential
arc are instantiated enough, we connect them with an arc. However, we will not
do this when we do not know that the arguments are instantiated enough, because
we want to be sure that there cannot be an infinite path consisting of arcs.
Consider for example the program

\begin{eqnarray*}
&& \tt int(0).\\
&& \tt int(s(X)) \imp int(X).
\end{eqnarray*}
with the query {\tt int(Y)}
and the term-size norm. From the rule we get the weighted rule graph that
is shown in Figure~\ref{int_wrg},
but as Figure~\ref{int_inf_deriv} shows there is an infinite derivation.
\wgefig{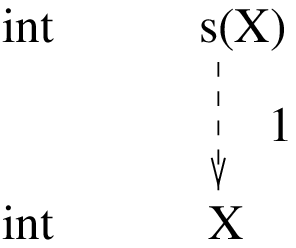}{Weighted rule graph}{int_wrg}{0.15\textwidth}
\eat{ 
\setlength{\unitlength}{1.1mm}
\begin{picture}(100,45)(5,10)
\put(10,40){int}
\put(40,40){s(X)}
\put(10,20){int}
\put(40,20){X}
\put(42,38){\line(0,-1){3}}
\put(42,33.5){\line(0,-1){3}}
\put(42,29){\vector(0,-1){4}}
\put(44,31){1}
\end{picture}
\noindent 
} 
\begin{figure}[htb]
\setlength{\unitlength}{1mm}
\begin{picture}(100,40)
\put(0,35){$\leftarrow int(Y)$}
\put(15,30){$\{Y\mapsto s(Y1)\}$}
\put(0,25){$\leftarrow int(Y1)$}
\put(15,20){$\{Y1\mapsto s(Y2)\}$}
\put(0,15){$\leftarrow int(Y2)$}
\put(15,7){$\vdots$}
\end{picture}
\caption{Infinite derivation}
\label{int_inf_deriv}
\end{figure}

\eat{ 
\begin{picture}(100,40)
\put(0,35){$\leftarrow int(Y)$}
\put(15,30){$\{Y\mapsto s(Y1)\}$}
\put(0,25){$\leftarrow int(Y1)$}
\put(15,20){$\{Y1\mapsto s(Y2)\}$}
\put(0,15){$\leftarrow int(Y2)$}
\put(15,7){$\vdots$}
\end{picture}
} 

We return now to the original example. We have the query 
\[mult(s(s(0)),s(0),Z)\] which we abstract to the query pattern
$mult(b,b,f)$ with empty constraint list. We now construct the summaries
of the augmented argument mappings associated with it, terms that will be
defined presently. Our definition of augmented argument mapping
differs from that in \cite{Sagiv}, for reasons that will be
explained.
The basic idea is that  we take a rule $r$, for which the head predicate
is the same as in the query, and a subgoal  $s$ in the body of $r$,
and try to approximate the `real' query-mapping pair corresponding to the
head of $r$ and $s$ in the LD-tree.
 We do this by using information we have from the weighted
rule graph of $r$ and also, if we have them, results of {\bf instantiation analysis} and 
{\bf constraint inference} 
about the instantiations and constraints of the body subgoals preceding $s$,
which we assume have succeeded before we got to $s$. The method used for the instantiation analysis and constraint inference
is abstract interpretation. For the details see \cite{Lindenstrauss:Sagiv}.

\begin{definition}[Augmented Argument Mapping] An {\em augmented argument
mapping}, which is a mixed graph, is constructed for a rule $r$ and a subgoal $s$ in its body
as follows. 
\begin{itemize}
\item There are nodes for the argument  positions of the head of
$r$, for $s$, and for all subgoals that precede $s$. 
\item Nodes are blackened in
agreement with the rule and the instantiation analysis for subgoals
that precede $s$.\footnote{In~\cite{Sagiv} nodes that correspond to arguments that precede
$s$ are made black. This is justified there because of the assumption
that any variable in the head of a program clause also appears in its body, 
an assumption which causes all atoms in the success set of the program to be
 ground, but not in our more general setting. Another difference is the 
use of weighted arcs. For instance, if we have the configuration
with edges, arcs and weighted arcs, that is presented in 
Figure~\ref{configuration},
we can say that in the transitive closure there is an arc from $N_{1}$
to $N_{5}$, while we could not come to this conclusion if the weighted
arcs were ordinary arcs.} 
\item There are all the
edges and arcs that can be derived from the weighted rule graph and
from the  constraint inference for subgoals preceding $s$.
\item In the case of disjunctive information about constraints or instantiations
the augmented argument mapping will use one disjunct.
\item The graph is consistent.
\item The {\em domain} consists of the nodes corresponding to the head of $r$,
and the {\em range} consists of the nodes corresponding to $s$.
\end{itemize}
\end{definition}
\wgefig{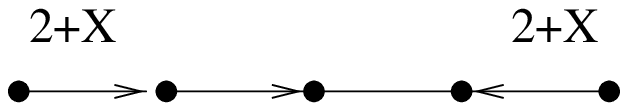}{The weighted arcs}{configuration}{0.40\textwidth}
A few words about the consistency: If, for instance, we are constructing
the augmented argument mapping for the rule
\begin{eqnarray*}
&& \tt a(X,Y) \imp b(X,Y), c(X,Y), d(X,Y).
\end{eqnarray*}
and the subgoal $d(X,Y)$ we cannot use for $b$ the constraint $gt(1,2)$
and for $c$ the conflicting constraint $gt(2,1)$, or for $b$ the instantiation
$b(ie,nie)$ and for $c$ the conflicting instantiation $c(nie,ie)$
(here $ie$
 denotes an argument that is instantiated enough, and $nie$ an argument that is not---note the difference between $ie$ and $nie$, which are mutually 
exclusive and $b$ and $f$, where the second possibility includes the first).
\eat{ 
\begin{picture}(105,20)(5,0)
\put(5,10){$\bullet$}
\put(30,10){$\bullet$}
\put(55,10){$\bullet$}
\put(80,10){$\bullet$}
\put(105,10){$\bullet$}
\put(12,15){$2+X$}
\put(91,15){$2+X$}
\put(5,5){$N_{1}$}
\put(30,5){$N_{2}$}
\put(55,5){$N_{3}$}
\put(80,5){$N_{4}$}
\put(105,5){$N_{5}$}
\put(5,11){\vector(1,0){23}}
\put(30,11){\vector(1,0){23}}
\put(55,11){\line(1,0){25}}
\put(105,11){\vector(-1,0){23}}
\end{picture}
}

\begin{definition}[Summary] If $\mu$ is a consistent augmented argument 
mapping,   then the {\em summary} of $\mu$ consists of the nodes in the 
domain and range of $\mu$ and the edges and arcs among these nodes (it is
undefined if $\mu$ is inconsistent).
\end{definition}

Summaries of augmented argument mappings give us approximations to `real'
mappings constructed for nodes that are direct offspring of each other.

We return now to the query $mult(b,b,f)$ and build the augmented argument
mappings derived for it from the rule

\begin{eqnarray*}
&& \tt mult(s(X),Y,Z)) \imp mult(X,Y,Z1), add(Z1,Y,Z).
\end{eqnarray*}

It should be noted that when we build the augmented argument mapping for a 
query we also take into account the instantiations and constraints of 
the query.

For the first subgoal we get the mapping presented in Figure~\ref{mapping1}.
\wgefig{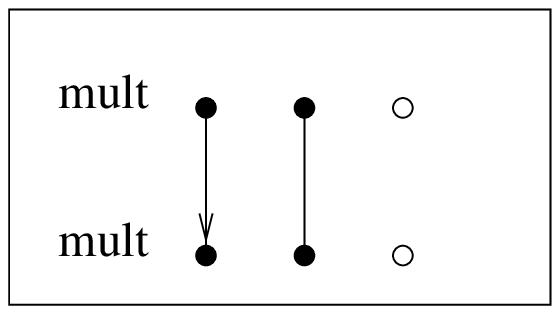}{Mapping for the first subgoal}{mapping1}{0.30\textwidth}
\eat{ 
\setlength{\unitlength}{0.8mm}
\fbox{
\begin{picture}(85,25)(50,-2)
\put(50,20){mult}
\put(50,0){mult}
\put(90,20){$\bullet$}
\put(110,20){$\bullet$}\put(50,20){mult}
\put(50,0){mult}
\put(90,20){$\bullet$}
\put(110,20){$\bullet$}
\put(130,20){$\circ$}
\put(90,0){$\bullet$}
\put(110,0){$\bullet$}
\put(130,0){$\circ$}
\put(91,20){\vector(0,-1){18}}
\put(111,20){\line(0,-1){20}}
\end{picture}
}
} 
\noindent The summary is identical to the mapping, and Figure~\ref{qm1}
presents the query-mapping pair obtained.
\wgefig{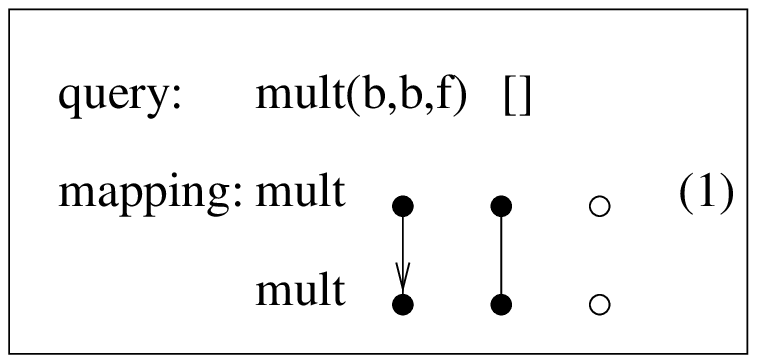}{Query-mapping pair for the first subgoal of {\tt mult}}{qm1}{0.40\textwidth}
\eat{ 
\setlength{\unitlength}{0.5mm}
\vspace{2mm}
\fbox{
\begin{picture}(170,50)(-30,-2)
\put(0,40){query:}
\put(50,40){mult(b,b,f)}
\put(110,40){[]}
\put(0,20){mapping:}
\put(50,20){mult}
\put(50,0){mult}
\put(90,20){\circle*{2}}
\put(110,20){\circle*{2}}
\put(130,20){\circle{3}}
\put(90,0){\circle*{2}}
\put(110,0){\circle*{2}}
\put(130,0){\circle{3}}
\put(-30,20){(1)}
\put(110,20){\line(0,-1){20}}

\put(90,19){\vector(0,-1){18}}

\end{picture}
}
\vspace{2mm}
} 

From the augmented argument mapping corresponding to the second subgoal of 
the rule 
we basically want to infer the relationship between $mult(s(X),Y,Z)$ and $add(Z1,Y,Z)$
assuming $mult(X,Y,Z1)$ has already been proved. This is where we use instantiation 
analysis and, possibly, constraint inference. If we did not use any information on
$mult(X,Y,Z1)$ we would get the augmented argument mapping presented in Figure~\ref{mapping2a}.
\wgefig{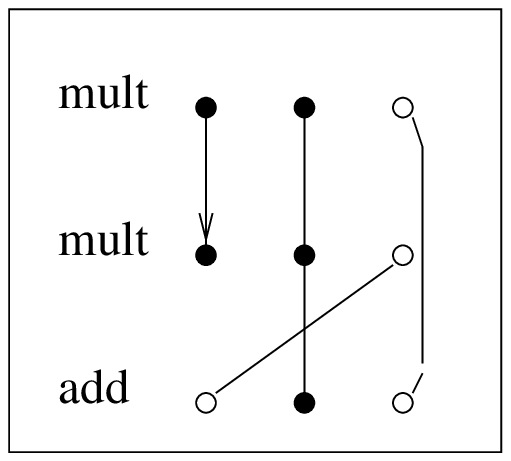}{Mapping for the second subgoal}{mapping2a}{0.30\textwidth}
\eat{ 
\fbox{
\begin{picture}(85,45)(50,-2)
\put(50,40){mult}
\put(90,40){$\bullet$}
\put(110,40){$\bullet$}
\put(130,40){$\circ$}
\put(50,20){mult}
\put(90,20){$\bullet$}
\put(110,20){$\bullet$}
\put(130,20){$\circ$}
\put(50,0){add}
\put(90,0){$\circ$}
\put(110,0){$\bullet$}
\put(130,0){$\circ$}

\put(91,40){\vector(0,-1){18}}
\put(111,40){\line(0,-1){20}}
\put(111,20){\line(0,-1){20}}
\put(128,19){\line(-2,-1){35}}
\put(132,21){\oval(5,40)[r]}
\end{picture}
}
} 
Now from the instantiation analysis we will get that the two possible instantiations for
$mult$ are $mult(ie,ie,ie)$ and $mult(ie,nie,ie)$ where $ie$ denotes a ground term and $nie$
denotes a non-ground term, i.e., a term that contains at least one variable. Since the
first two arguments of the intermediate subgoal are ground, so must the third one be. 
So we get the augmented argument mapping that is presented in Figure~\ref{mapping2b},
\wgefig{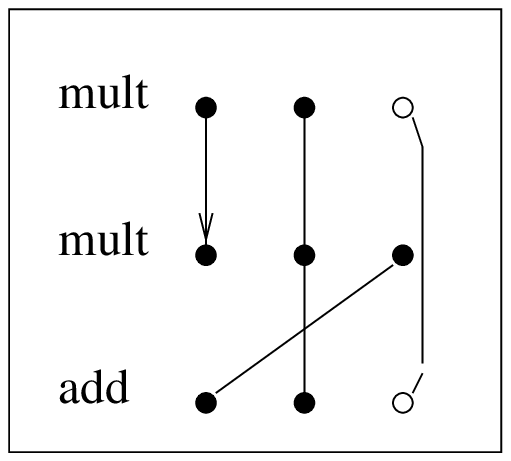}{Mapping for the second subgoal}{mapping2b}{0.30\textwidth}
\noindent
which gives rise to the query-mapping pair presented in Figure~\ref{qm2} 
\wgefig{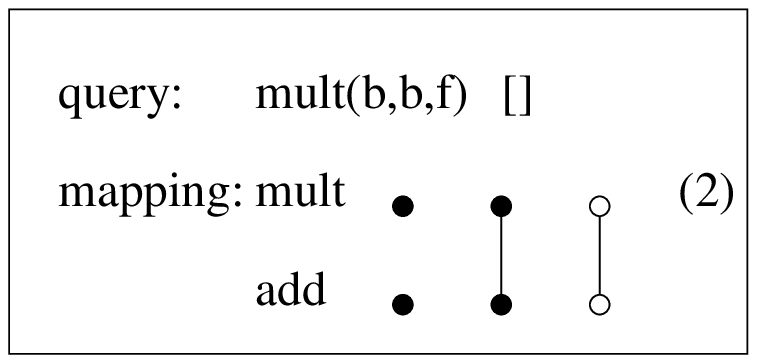}{Query-mapping pair for the second subgoal of {\tt mult}}{qm2}{0.40\textwidth}
and the new query $add(b,b,f)$ (with no constraints). If we had
not been able to use the results of the instantiation analysis we would have gotten the
query $add(f,b,f)$, which does not terminate, and our algorithm would just have said for 
the original query that there {\em may} be non-termination. 

Using the appropriate augmented argument mappings for the new query $add(b,b,f)$, we get the new query-mapping
pairs (Figures~\ref{qm3} and~\ref{qm4}) and no new queries.
\wgefig{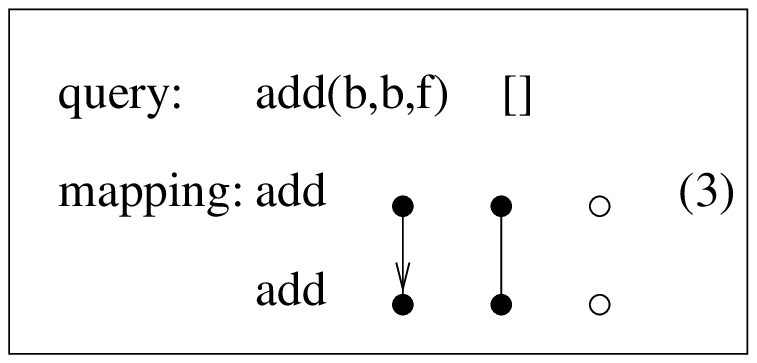}{Query-mapping pair for {\tt add}}{qm3}{0.40\textwidth}
\wgefig{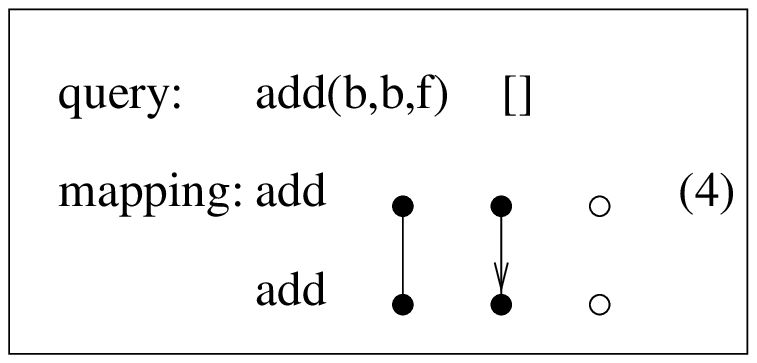}{Query-mapping pair for {\tt add}}{qm4}{0.40\textwidth}
\eat{ 
\setlength{\unitlength}{0.5mm}
\vspace{2mm}
\fbox{
\begin{picture}(170,50)(-30,-2)
\put(0,40){query:}
\put(50,40){add(b,b,f)}
\put(110,40){[]}
\put(0,20){mapping:}
\put(50,20){add}
\put(50,0){add}
\put(90,20){\circle*{2}}
\put(110,20){\circle*{2}}
\put(130,20){\circle{3}}
\put(90,0){\circle*{2}}
\put(110,0){\circle*{2}}
\put(130,0){\circle{3}}

\put(90,20){\vector(0,-1){19}}

\put(110,20){\line(0,-1){20}}
\put(-30,20){(3)}

\end{picture}
}
\vspace{2mm}
 
\noindent and 

\setlength{\unitlength}{0.5mm}
\vspace{2mm}
\fbox{
\begin{picture}(170,50)(-30,-2)
\put(0,40){query:}
\put(50,40){add(b,b,f)}
\put(110,40){[]}
\put(0,20){mapping:}
\put(50,20){add}
\put(50,0){add}
\put(90,20){\circle*{2}}
\put(110,20){\circle*{2}}
\put(130,20){\circle{3}}
\put(90,0){\circle*{2}}
\put(110,0){\circle*{2}}
\put(130,0){\circle{3}}

\put(90,20){\line(0,-1){20}}

\put(110,20){\vector(0,-1){19}}
\put(-30,20){(4)}

\end{picture}
}
\vspace{2mm}
} 

Now we have to apply composition to the query-mapping pairs we have created thus far. Recall the
following definitions from \cite{Sagiv}:

\begin{definition}[Composition of Mappings]
 If the range of a mapping $\mu$ and the domain of a mapping
$\nu$ are labeled by the same predicate, then the {\em composition of the
mappings} $\mu$ and $\nu$, denoted $\mu \circ \nu$, is obtained by unifying
each node in the range of $\mu$ with the corresponding node in the domain
of $\nu$. When unifying  two nodes, the result is a black node if at least
one of the nodes is black, otherwise it is a white node. If a node becomes 
black, so do all nodes connected to it with an edge. The domain of 
$\mu \circ \nu$ is that of $\mu$, and its range is that of $\nu$. The edges 
and arcs of 
$\mu \circ \nu$ consist of the transitive closure of the union of the edges and
 arcs of $\mu$ and $\nu$.
\end{definition}

\begin{definition}[Composition of Query-Mapping Pairs]
 Let $(\pi _{1},\mu _{1})$ and 
$(\pi _{2},\mu _{2})$ be query-mapping pairs, such that the range of $\mu _{1}$ is
identical to $\pi _{2}$. The composition of $(\pi _{1},\mu _{1})$ and 
$(\pi _{2},\mu _{2})$ is $(\pi _{1},\mu)$, where $\mu$ is the summary of $\mu _{1} \circ \mu_{2}$
(and, hence, the composition is undefined if $\mu _{1} \circ \mu_{2}$ is inconsistent).
\end{definition}

By repeatedly composing the approximations we got thus far we get the following
new pairs.

Composition of pairs (1) and (2) gives a new pair (5); pairs (3) and (4)
give a new pair (6); (2) and (6) give (7). These new pairs
are presented by Figures~\ref{qm5},\ \ref{qm6} and \ref{qm7} respectively.
\wgefig{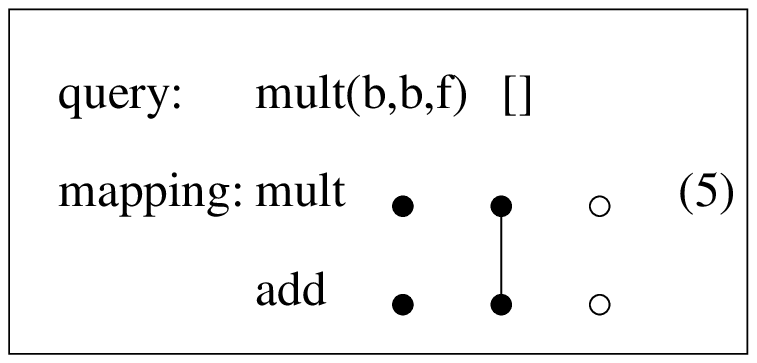}{Query-mapping pair for {\tt add}}{qm5}{0.40\textwidth}
\wgefig{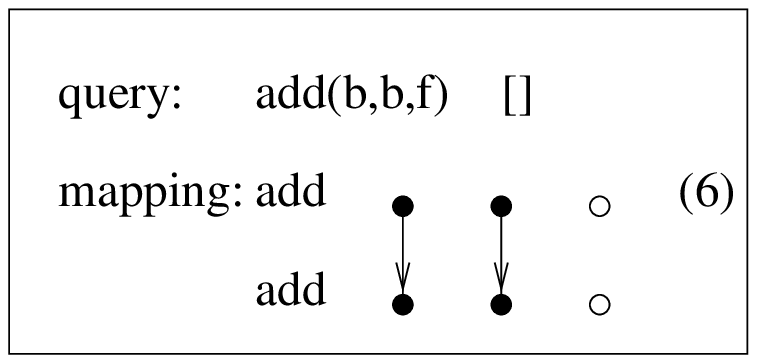}{Query-mapping pair for {\tt add}}{qm6}{0.40\textwidth}
\wgefig{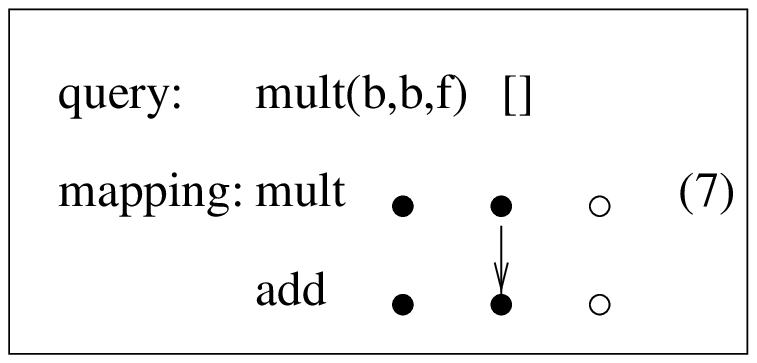}{Query-mapping pair for {\tt add}}{qm7}{0.40\textwidth}
\eat{ 
\setlength{\unitlength}{0.5mm}
\vspace{2mm}
\fbox{
\begin{picture}(170,50)(-30,-2)
\put(0,40){query:}
\put(50,40){mult(b,b,f)}
\put(110,40){[]}
\put(0,20){mapping:}
\put(50,20){mult}
\put(50,0){add}
\put(90,20){\circle*{2}}
\put(110,20){\circle*{2}}
\put(130,20){\circle{3}}
\put(90,0){\circle*{2}}
\put(110,0){\circle*{2}}
\put(130,0){\circle{3}}
\put(110,20){\line(0,-1){20}}
\put(-30,20){(5)}

\end{picture}
}
\vspace{2mm}

Composition of pairs (3) and (4):

\setlength{\unitlength}{0.5mm}
\vspace{2mm}
\fbox{
\begin{picture}(170,50)(-30,-2)
\put(0,40){query:}
\put(50,40){add(b,b,f)}
\put(110,40){[]}
\put(0,20){mapping:}
\put(50,20){add}
\put(50,0){add}
\put(90,20){\circle*{2}}
\put(110,20){\circle*{2}}
\put(130,20){\circle{3}}
\put(90,0){\circle*{2}}
\put(110,0){\circle*{2}}
\put(130,0){\circle{3}}

\put(90,19){\vector(0,-1){18}}

\put(110,19){\vector(0,-1){18}}
\put(-30,20){(6)}

\end{picture}
}

Composition of pairs (2) and (6):

\vspace{2mm}

\setlength{\unitlength}{0.5mm}
\vspace{2mm}
\fbox{
\begin{picture}(170,50)(-30,-2)
\put(0,40){query:}
\put(50,40){mult(b,b,f)}
\put(110,40){[]}
\put(0,20){mapping:}
\put(50,20){mult}
\put(50,0){add}
\put(90,20){\circle*{2}}
\put(110,20){\circle*{2}}
\put(130,20){\circle{3}}
\put(90,0){\circle*{2}}
\put(110,0){\circle*{2}}
\put(130,0){\circle{3}}

\put(110,19){\vector(0,-1){18}}
\put(-30,20){(7)}
\end{picture}
}
\vspace{2mm}
} 
No more query-mapping pairs can be created. 

Since for each of the above query-mapping pairs, if the circular variant exists it has a forward 
positive cycle, it follows, since for each call branch its image under
 $\alpha _{app}$ subsumes its image under $\alpha_{real}$,  that queries of the form $mult(b,b,f)$
terminate.

If in the above example we would have used  results of the constraint inference for the one 
intermediate subgoal we had, we would have gotten more query-mapping pairs,
but again every circular variant would have had a forward positive cycle, so we
would have been able to show termination.
However, if it is possible to prove termination without constraint
inference, there is no reason to use it, because the more query-mapping pairs 
there are the longer the termination proof takes.

The following theorem holds:
\begin{theorem}\label{qm-th}
Let the LD-tree for a query and program and a symbolic linear norm be given.
Define the structured shadow $\alpha _{app}$ as above. If all the circular
variants that can be created from query-mapping pairs in the image of
$\alpha _{app}$ have a forward positive cycle then the tree must be finite,
i.e., there is termination for the query with Prolog's computation rule.
\end{theorem}
This theorem follows from Theorem~\ref{LS-real-termination-criterion} 
if we notice that for every call
branch $B$ we have that $\alpha _{real}(B)$ is subsumed by
$\alpha _{app}(B)$.

Actually, since we have here a structured shadow, the element whose existence is proved in the basic lemma is both circular and idempotent. So 
we can formulate a stronger theorem, which is more efficient to implement:
\begin{theorem}\label{qm-th-stronger}
Let the LD-tree for a query and program and a symbolic linear norm be given.
Define the structured shadow $\alpha _{app}$ as above. If all the circular
idempotent  query-mapping pairs in the image of
$\alpha _{app}$ have an arc from an argument in the domain to the corresponding
argument in the range, then the tree must be finite,
i.e., there is termination for the query with Prolog's computation rule.
\end{theorem}
It is not difficult to see that the condition of Theorem~\ref{qm-th} implies the 
condition of Theorem~\ref{qm-th-stronger}, since if we have a circular idempotent pair
for which the circular variant has a forward positive cycle, and compose it 
with itself the right number of times (that is, the number of circular arcs on
the forward positive cycle), we get an arc from an argument in the 
domain to the corresponding argument in the range.

Theorem~\ref{qm-th-stronger} is really stronger than Theorem~\ref{qm-th}, as
the following program shows:
\begin{eqnarray*}
&& \tt p(0,0).\\
&& \tt p(s(X),Y) \imp p(f(0),X).\\
&& \tt p(X,s(Y)) \imp p(Y,f(0)).
\end{eqnarray*}
In this case there is termination for queries of the form $p(b,b)$, but, using
the term-size norm, this can only be deduced from Theorem~\ref{qm-th-stronger}
and not Theorem~\ref{qm-th}, because there are circular variants without
forward positive cycle.

It is interesting to note that the rule of composition does not hold for
$\alpha _{real}$, so it is not a structured shadow. For instance if we take the program

\begin{eqnarray*}
&& \tt p(X,a) \imp q(X).\\
&& \tt q(X) \imp r(X,a).
\end{eqnarray*}

\noindent
and the LD-tree for {\tt p(X,a)}
\begin{eqnarray*}
(1) & \leftarrow & p(X,a)\\
(2) & \leftarrow & q(X)\\
(3) & \leftarrow & r(X,a)
\end{eqnarray*}
and the term-size norm, then
the image under $\alpha _{real}$ of the branch from (1) to (2) composed
with the image under $\alpha _{real}$ of the branch from (2) to (3)
will not contain the edge between the second arguments of $p$ and $r$
that is in the image under $\alpha _{real}$ of the branch between (1)
and (3). We will always have that
$\alpha _{real}(B_{1})*\alpha _{real}(B_{2})$ subsumes
$\alpha _{real}(B_{1}*B_{2})$.

The following optimization  of Theorem~\ref{qm-th} holds: It is enough to
consider only query-mapping pairs in which the predicate of the domain
and the predicate of the range are in the same strongly connected component
of the predicate dependency graph. 

\section{Logic programs containing arithmetic predicates}
\label{Preliminaries}

The algorithm we describe next would come into play
only when the usual termination analyzers fail to prove
termination using the structural arguments of predicates. 
As a first step it verifies the presence of an integer loop in 
the program. If no integer  loop is found, the possibility of 
non-termination is reported, meaning that the termination
cannot be proved by this technique. If integer loops are found, each of them 
is taken into consideration. 
The algorithm starts by discovering integer positions in the program, 
proceeds with creating appropriate abstractions, based on the integer loops,
and concludes by applying an extension of the query-mapping pairs technique. 
The formal algorithm is presented in Subsection~\ref{Algorithm}.

The structured shadow we define in this case assigns to the branch from
node $\leftarrow r_1,\ldots,r_n$ to its direct offspring 
$\leftarrow s_1,\ldots, s_k$, where $\theta$ is the composition of the
substitutions between the nodes, the following query-mapping pair:
each atom is abstracted to a pair $(\mbox{\sl predicate}, 
\mbox{\sl constraint})$, where the constraint is one from a finite set of 
mutually exclusive numerical constraints (for example, $\mbox{\sl arg1}>0,
\mbox{\sl arg1}>\mbox{\sl arg2}$, where {\sl arg1} and {\sl arg2} are
respectively the first and the second arguments of the atom).
The query is the abstraction of $r_1$. The mapping of the query-mapping pair,
is as before, a quadruple---the domain, the range, edges and arcs.
The domain of the query-mapping pair is the abstraction of $r_1\theta$, the 
range is the abstraction of $s_1$, and there are edges and arcs between 
nodes of $r_1\theta$ and $s_1$. Edges and arcs correspond to numerical 
equalities and inequalities of the respective arguments. When composing two
query-mapping pairs numerical nodes are unified only if they have the same 
constraint (remember that the constraints are mutually exclusive). Termination
is shown by means of a non-negative termination function of the arguments of 
an atom, that decreases from the domain to the range (cf.\ \cite{Floyd}). 
Note that in the 
numerical part of the program we will use both the query-mapping pairs
relative to the norm and the new kind of numerical query-mapping pairs.
As we will see later on, sometimes in order
to prove termination (cf. Example~\ref{both:types:of:pairs})
both kinds of query-mapping pairs are essential.

The technique we present in this section allows us to analyze correctly 
on the one hand common examples of Prolog programs
(such as 
{\sl factorial}~\cite{Debray},
{\sl Fibonacci, Hanoi}~\cite{Maria:Benchmarks},
{\sl odd\_even}~\cite{Plumer:Book}, {\sl between}~\cite{Apt:Book},
{\sl Ackermann}~\cite{Sterling:Shapiro}),
and on the other hand more
difficult examples, such as {\sl gcd} and {\sl mc\_carthy\_91}~\cite{Manna:McCarthy,Knuth,Giesl}. Note that some of these examples were previously
considered in the literature on termination. However, they were always 
assumed to be given in the {\em successor notation\/}, thus solving the
problem of well-foundedness. Moreover, the analysis of 
some of these examples, such as
{\sl gcd}, required special techniques~\cite{Lindenstrauss:Sagiv:Serebrenik:L}.

\subsection{The 91 function}
\label{Motivation}
We start by illustrating informally the use of our algorithm for proving the termination
of the 91 function. This convoluted function was invented by John McCarthy for 
exploring properties of recursive programs, and is considered to be a good
test case for automatic verification systems (cf. \cite{Manna:McCarthy,Knuth,Giesl}). The treatment here is on the intuitive level.
Formal details will be given in subsequent sections.

Consider the clauses:

\begin{example}
\label{91:function}
\begin{eqnarray*}
&&\tt \mc(X,Y) \imp X>100, Y\;\;is\;\;X-10.\\
&&\tt \mc(X,Y) \imp X\leq 100, Z\;\;is\;\;X+11, \mc(Z,Z1),\\
&& \hspace{3.5cm}\tt \mc(Z1,Y).
\end{eqnarray*}

\noindent
and assume that a query of the form {\tt \mc($i$,$f$)} is given, 
that is, a query in which
the first argument is bound to an integer, and the second is free. This
program computes the same answers as the following one:
\begin{eqnarray*}
&&\tt \mc(X,Y) \imp X>100, Y\;\;is\;\;X-10.\\
&&\tt \mc(X,91) \imp X\leq 100.
\end{eqnarray*}
\noindent
with the same query. Note, however, that while the termination of the latter
program is obvious, since there is no recursion in it, the termination of
the first one is far from being trivial and a lot of effort was dedicated
to find termination proofs for it (\cite{Manna:McCarthy,Knuth,Giesl}).
\end{example}

Our algorithm starts off by {\bf discovering numerical arguments}. This step
is based on abstract interpretation, 
and as a result both arguments of {\tt \mc} are proven to be numerical. 
Moreover,
they are proven to be of integer type. The importance of knowledge of this 
kind and 
techniques for its discovery are discussed in Subsection~\ref{Discovering}.

The next step of the algorithm is the inference of the (finite) {\bf integer 
abstraction domain} which will help overcome difficulties caused by the fact that the
(positive and negative) integers with the usual (greater-than or less-than) order are not well-founded. Integer abstractions are
derived from arithmetic comparisons in the bodies of rules. However,
a simplistic approach may be insufficient and the more powerful techniques
presented in Section~\ref{IntDomain} are sometimes essential. In our case 
the domain \[\{(-\infty,89], [90,100], [101,\infty)\}\] of intervals is deduced. For the sake of
convenience we denote this tripartite domain by $\{${\sl small, med, big}$\}$.

In the next step, we {\bf use abstract interpretation to describe answers to 
queries}. This allows us to infer numerical inter-argument relations of a 
novel type.  
In Section~\ref{AbstrInterp} the technique for inference of constraints of this
kind is presented. For our running example we get the following abstract atoms:

\medskip
\begin{tabular}{ll}
{\tt \mc({\sl big},{\sl big})}& {\tt \mc({\sl med},{\sl med})}\\
{\tt \mc({\sl big},{\sl med})}& {\tt \mc({\sl small},{\sl med})}
\end{tabular}
\medskip

\noindent These abstract atoms characterize the answers of the program.
 
The concluding step creates {\bf query-mapping pairs} in the fashion of~\cite{Lindenstrauss:Sagiv}. This process uses the abstract descriptions
of answers to queries and is described in Section~\ref{QM}.
In our case, we obtain among others, the query-mapping pair
having the query {\tt \mc($i$,$f$)}, where $i$ denotes an integer argument and
$f$ an unrestricted one, and the mapping presented in 
Figure~\ref{qm91a}. The upper nodes correspond to argument positions of the
head of the recursive clause, 
and the lower nodes---to argument positions of the second
recursive subgoal in the body. 
Circled black nodes denote integer argument positions,
and white nodes denote positions
on which no assumption is made. The 
arc denotes an increase of the first argument, in the sense that
the first argument in the head is less than 
the first argument in the second recursive subgoal. 
Each set of nodes is accompanied by a set of constraints. Some
could be  inter-argument relations of the type considered
in~\cite{Lindenstrauss:Sagiv}. In our example this subset is empty. The rest
are constraints based on the integer abstraction domain. In this
case, that set contains the constraint that the first argument is in
{\sl med}. 
\wgefig{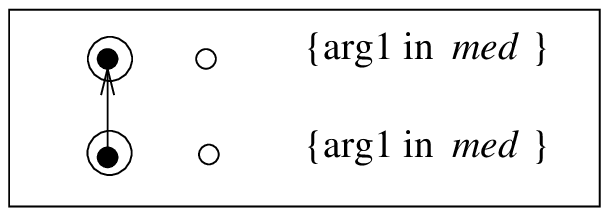}{Mapping for McCarthy's 91 function}{qm91a}{0.40\textwidth}
The query-mapping pair presented is circular (upper and lower nodes are the same),
but the termination tests of~\cite{Lindenstrauss:Sagiv,Codish:Taboch} fail.  
Thus, a termination function must be guessed.
For this loop we can use the function $100-${\sl arg1}, where 
{\sl arg1} denotes the first argument of the atom.
The value of this function decreases while traversing the given 
query-mapping pair from the upper to the lower nodes.
Since it is also bounded from below ($100\geq ${\sl arg1}),
this query-mapping pair may be traversed only finitely many times. The
same holds for the other circular query-mapping pair in this case. 
Thus, termination is proved.


\subsection{Arithmetic Loops}
We start our discussion on termination of numerical computations
by providing a formal definition of numerical loop, analyzing the
problems one discovers when reasoning about termination of numerical loops
and explain  why we restrict ourselves to integer loops.
In the end of this section we discuss a technique for discovering numerical
argument positions that we'll base our termination analysis on.

\subsubsection{Numerical and integer loops}
Our notion of numerical loop is based on the predicate dependency graph
(cf.\ \cite{Plumer:Book}):

\begin{definition}
Let $P$ be a program and let $\Pi$ be a strongly connected component
in its predicate dependency graph.  Let $S\subseteq P$ be the set of 
program clauses, associated with $\Pi$ (i.e. those clauses that have the 
predicates of $\Pi$ in their head).
$S$ is called {\em loop} if there is a cycle through predicates of $\Pi$.
\end{definition}
\begin{definition}
A loop $S$ is called {\em numerical\/} if there is a clause \[H\imp B_1,\ldots,
B_n\] in $S$, such that for some $i$, 
$B_i\equiv \mbox{{\sl Var\ }{\tt is\ }{\sl Exp}}$,
and either {\sl Var} is equal to some argument of $H$ or {\sl Exp} is an arithmetic expression
involving a variable that is equal to some argument of $H$.
\end{definition}

However,
termination of numerical loops that involve numbers that are not integers
often depends on the specifics of
implementation and hardware, so
 we  limit ourselves
to ``integer loops'', rather than all numerical loops. 
The following examples illustrate actual behavior that
contradicts intuition---a loop
that should not terminate terminates, while a loop that 
should terminate does not. We checked the
behavior of these examples on UNIX, with 
the CLP(Q,R) library~\cite{CLP:Manual} of SICStus Prolog~\cite{SICStus:Manual},
CLP(R)~\cite{Jaffar:Maher} and XSB~\cite{XSB:Manual}.

\begin{example}
\label{ex:real:loops}
Consider the following program. The goal {\tt p(1.0)} terminates although we
would expect it not to terminate. On the other hand the goal {\tt  q(1.0)}
does not terminate, although we would expect it to terminate.
\begin{eqnarray*}
&& \tt p(0.0)\imp !.\\
&& \tt p(X) \imp X1\;\;is\;\;X/2,\;\;p(X1).\\
&&\\
&& \tt q(0.0)\imp !.\\
&& \tt q(X) \imp X1\;\;is\;\;X\;-\;0.1,\;\;q(X1).
\end{eqnarray*}
\end{example}

One may suggest that assuming that the program does not contain division
and non-integer
constants will solve the problem. The following example shows that this
is not the case:
\begin{example}
\label{ex:num:loop}
\begin{eqnarray*}
&& \tt r(0).\\
&& \tt r(X) \imp X>0,\;\;X1\;\;is\;\;X-1,\;\;r(X1).
\end{eqnarray*}
The predicate ${\tt r}$ may be called with a
real, non-integer argument, and then its behavior is implementation dependent.
For example, one would expect that ${\tt r(0.0)}$ will succeed and
${\tt r(0.000000001)}$ will fail. However, in SICStus Prolog both goals
fail, while in CLP(R) both of them succeed!
\end{example}

Therefore, we limit ourselves to integer loops, that is numerical loops 
involving integer constants and arithmetical calculations over integers:
\begin{definition}
A program $P$ is {\em integer-based} if, given a query such that all 
numbers appearing in it are integers, all subqueries that arise have this 
property as well.
\end{definition}
\begin{definition}
A numerical loop $S$ in a program $P$ is called an {\em integer\/} loop if $P$
is integer-based.
\end{definition}

Termination of a query may depend on whether its argument is an integer, as
the following example shows:

\begin{verbatim}
p(0).
p(N) :- N > 0, N1 is N - 1, p(N1).
p(a) :- p(a).
\end{verbatim}

\noindent For this program, $p(X)$ for integer $X$ terminates, while $p(a)$ does not. 

So we extend our notion of query pattern.
Till now a  query pattern was an atom with the
same predicate and arity as the query, and arguments  $b$ 
(denoting an argument that is instantiated enough with respect to the norm) 
or $f$ (denoting an argument on which no 
assumptions are made). Here, we extend the notion to include 
arguments of the form $i$,
denoting an argument that is an integer (or integer expression). 
Note
that $b$ includes the possibility of $i$ in the same way that $f$ includes the 
possibility $b$. 
In the diagrams to follow we denote $i$-arguments by 
circled black nodes, and as before, $b$-arguments by black nodes and $f$-arguments by white nodes.

Our termination analysis is always performed with respect to a given program and 
a query pattern. A positive response guarantees termination of every query that matches the 
pattern.  

\subsubsection{Discovering integer arguments}
\label{Discovering}
Our analysis that will be discussed in the subsequent sections is based
on the size relationships between ``integer arguments''. So we have to 
discover which arguments are integer arguments. In simple programs this is immediate,
but there may be more complicated cases.

The inference of integer arguments is done in two phases---bottom-up and top-down. 
Bottom-up inference is similar to type analysis 
(cf.\ \cite{Codish:Lagoon,Boye:Maluszynski}), 
using the abstract domain
$\{\mbox{{\sl int}},\mbox{{\sl not\_int}}\}$ 
and the observation that an argument 
may became {\sl int} only
if it is obtained from {\tt is/2} or is bound to an integer expression of 
arguments 
already found to be {\sl int} (i.e. the abstraction of 
$\mbox{{\sl int}}+\mbox{{\sl int}}$
is {\sl int}). 
Top-down inference is query driven and is similar to the ``blackening'' 
process, 
described in~\cite{Lindenstrauss:Sagiv}, 
only in this case the information propagated is being an integer expression 
instead of ``instantiated enough''. 

Take for example the program
\begin{verbatim}
p(0).
p(N) :- N > 0, N1 is N - 1, p(N1).
p(a) :- p(a).
q(X) :- r(X,Y), p(Y).
r(b,a).
r(X,X).
\end{verbatim}
Denoting by $int$ an integer argument, by $gni$ an argument that is ground but
 not integer, and by $ng$ an argument that is not ground, we get from bottom-up 
instantiation analysis that the only pattern possible for $r(X,Y)$ atoms that
are logical consequences of the program are \[r(int,int),r(gni,gni),r(ng,ng)\]
Now we get from top down analysis that a query $q(i)$ gives rise to the
query $p(i)$ and hence terminates.

The efficiency of discovering numerical arguments may be improved
by a preliminary step of guessing the numerical argument positions. The
guessing is based on the knowledge that variables appearing in comparisons
or {\tt is/2}-atoms should be numerical. Instead of considering the whole 
program it is sufficient in this case to consider only clauses of
predicates having clauses with the guessed arguments
and clauses of predicates on which they depend. The 
guessing as a preliminary step becomes crucial when considering
``real-world'' programs that are large, while their numerical part is 
usually small.

\subsection{Integer Abstraction Domain}
\label{IntDomain}
In this subsection we present a technique that allows us to overcome the 
difficulties caused by the fact that
the integers with the usual order are not 
well-founded. Given a program $P$ we introduce a finite abstraction domain,
representing integers. The integer abstractions are derived from the subgoals
involving integer arithmetic positions. 

Let $S$ be a set of clauses in $P$, consisting of an integer loop and 
all the clauses for predicates on which the predicates of the integer loop 
depend. As a first step for definining the abstract domain for each 
recursive predicate $p$ in $S$ we
obtain the set of comparisons ${\cal C}_p$. If $p$ is clear from the
context we omit the index.

More formally, we consider as a 
{\em comparison}, an atom of the form $t_1 \rho t_2$, such that 
$t_1$ and $t_2$ are either variables or constants and 
$\rho\in \{<,\leq,\geq,>\}$. Our aim in restricting ourselves
 to these atoms is to ensure the
finiteness of ${\cal C}$. Other decisions can be made as long as finiteness
is ensured. Note that by excluding $\not =$ and $=$ we
do not limit the generality of the analysis. Indeed if $t_1 \not = t_2$
appears in a clause it may be replaced by two clauses containing  
$t_1 > t_2$ and $t_1 < t_2$ instead of $t_1 \not = t_2$, 
respectively. Similarly, if the clause contains a subgoal $t_1 = t_2$, 
the subgoal may be replaced by two subgoals $t_1 \geq t_2, t_1 \leq t_2$.
Thus, the equalities we use in the examples to follow should be seen as 
a brief notation as above.

In the following subsections we present a number of techniques 
to infer ${\cal C}$ from the clauses of $S$.

We define ${\cal D}_p$ as the set of pairs $(p,c)$, for all satisfiable $c \in 2^{{\cal C}_p}$. Here we interpret $c \in 2^{{\cal C}_p}$ as a conjunction
 of the comparisons in $c$ and the negations of the comparisons in
${\cal C}_p \setminus c$.
The abstraction domain ${\cal D}$ is taken as the union of the sets
${\cal D}_p$ for the recursive predicates $p$ in $S$.
Simplifying the domain may improve the running time of the analysis, however
it may make it less precise. 

\subsubsection{The simple case---collecting comparisons}

The simplest way to obtain ${\cal C}$ from the clauses of $S$ is
to consider the comparisons appearing in the bodies of recursive clauses
and restricting integer positions in their heads (we limit ourselves to the
recursive clauses, since these are the clauses that can give rise to circular pairs).

We would like to view ${\cal C}$ as a set of comparisons of head argument
positions. Therefore we assume in the simple case that $S$ is
{\em partially normalized}, that is, all head {\em integer\/} 
argument positions in clauses of $S$ are occupied by distinct variables. 
This assumption holds for all the examples considered so far.
This assumption will not be necessary with the more powerful technique 
presented in the next subsection.

\begin{example}
\label{comparisons}

Consider
\begin{eqnarray*}
&&\tt t(X)\imp X>5, X<8, X<2, X1\;is\;X+1, X1<5, t(X1).
\end{eqnarray*}
Let {\tt t($i$)} be a query pattern for the program above. 
In this case, the first argument of {\tt t} is an integer argument.
Since {\tt X1} does not appear in the head of the first clause
{\tt X1<5} is not considered and,
thus, ${\cal C}= \{X>5,X<8,X<2\}$. We have in this example
only one predicate and the union is over the singleton set.
So, ${\cal D} = \{X<2,2\leq X\leq 5,5<X<8,X\geq 8\}$.
\end{example}

The following example evaluates the {\sl mod\/} function. 
\begin{example}
\label{mod}
\begin{eqnarray*}
&& \tt mod(A,B,C) \imp A\geq B, B>0, D\;is\;A-B, mod(D,B,C).\\
&& \tt mod(A,B,C) \imp A< B, A \geq 0, A = C.
\end{eqnarray*}
Here we ignore the second clause since it is not recursive. Thus, by 
collecting 
comparisons from the first clause, ${\cal C} _{\mbox{\tt mod}}= \{
arg1\geq arg2, arg2>0\}$ and thus,
by taking all the conjunctions of comparisons of ${\cal C}$
and their negations, we obtain ${\cal D}_{\mbox{\tt mod}} = 
\{(\mbox{\tt mod},\mbox{\sl arg1}\geq \mbox{\sl arg2} \;\&\; \mbox{\sl arg2}>0),
(\mbox{\tt mod},\mbox{\sl arg1}\geq \mbox{\sl arg2} \;\&\; \mbox{\sl arg2}\leq 0),
(\mbox{\tt mod},\mbox{\sl arg1}< \mbox{\sl arg2} \;\&\; \mbox{\sl arg2}>0),
(\mbox{\tt mod},\mbox{\sl arg1}< \mbox{\sl arg2} \;\&\; \mbox{\sl arg2}\leq 0)\}$.
\end{example}

However, sometimes the abstract domain obtained in this way is insufficient
for proving termination, and thus, should be refined. The domain may be
refined by enriching the underlying set of comparisons. Possible
ways to do this are using inference of comparisons instead of collecting them, 
or performing an unfolding, and applying the collecting or inference 
techniques to the unfolded program. 

\subsubsection{Inference of Comparisons}

As mentioned above, sometimes the abstraction domain obtained from 
comparisons appearing in $S$ is insufficient. 
Instead of collecting comparisons, appearing in bodies of clauses,
we collect certain comparisons that are {\em implied} by bodies of clauses.
For example, {\tt X is Y+Z} implies the constraint
{\sl X=Y+Z} and {\tt functor(Term,Name,Arity)} implies {\sl Arity}$\geq 0$. 

As before, we restrict ourselves to recursive clauses and comparisons that
constrain integer argument positions of heads. 
Since a comparison that is contained in the body is implied by it, 
we always get
a superset of the comparisons obtained by the collecting technique, presented 
previously. The set of comparisons inferred depends on the power of the 
inference engine used (e.g. CLP-techniques may be used for this purpose). 

We define the abstract domain ${\cal D}$ as above. Thus, the 
granularity of the abstract domain also depends on the power of the inference
engine.

\subsubsection{Unfolding}

Unfolding 
(cf.~\cite{Tamaki:Sato,Bossi:Cocco,Apt:Book,Lindenstrauss:Sagiv:Serebrenik:L}) allows us to generate
a sequence of abstract domains, such that each refines the previous.

More formally, let $P$ be a program and let 
$H\imp B_1,\ldots,B_n$ be a recursive rule in $P$. 
Let $P_1$ be the result of unfolding an atom $B_i$ in 
$H\imp B_1,\ldots,B_n$ in $P$. Let $S_1$ be a set of clauses in $P_1$, 
consisting of an integer loop and the clauses 
of the predicates on which the integer loop predicates depend.
 
Obtain ${\cal D}$ for the clauses of $S_1$ 
either by collecting comparisons from
rule bodies or by inferring them, and use it as a new abstraction domain
for the original program.
If the algorithm still fails to prove termination, the process of
unfolding can be repeated.

\begin{example}
\label{domain_for_mc}
Unfolding {\tt \mc(Z1,Z2)} in the recursive clause we obtain a new program
for the query {\tt \mc($i$,$f$)}
\begin{eqnarray*}
&& \tt \mc(X,Y) \imp X>100,\;Y\;\;is\;\;X-10.\\
&& \tt \mc(X,Y) \imp X\leq 100,\;Z1\;\;is\;\;X+11, Z1>100,\\
&& \tt \hspace{1.5in} \;Z2\;\;is\;\;Z1-10,\mc(Z2,Y).\\
&& \tt \mc(X,Y) \imp X\leq 100,\;Z1\;\;is\;\;X+11, Z1\leq 100,\;Z3\;\;is\;\;Z1+11,\\
&& \tt \hspace{1.6in}\mc(Z3,Z4),\mc(Z4,Z2),\\
&& \tt \hspace{1.6in}\mc(Z2,Y). 
\end{eqnarray*}
Now if we use an inference engine that is able to discover that {\tt X is Y+Z}
implies the constraint {\sl X=Y+Z}, we obtain the
following constraints on the bound head integer variable $X$
(for convenience we omit redundant ones):
From the second clause we obtain:
$X\leq 100$, and since $X+11>100$ we get $X>89$. Similarly, from the third
clause: $X\leq 89$. Thus, ${\cal C}=\{X\leq 89,X>89 \wedge X\leq 100\}$
Substituting this in the definition of ${\cal D}$, and removing inconsistencies
and redundancies, we obtain ${\cal D}=\{X\leq 89,X>89 \wedge X\leq 100, 
X>100\}$.
\end{example}

\subsubsection{Propagating domains}
\label{propagating}
The comparisons we obtain by the techniques presented 
above may restrict only {\sl some\/} subset of integer argument
positions. However, for the termination proof, information on integer arguments
outside of this subset may be needed. 
For example, as we will see shortly,
in order to analyze correctly {\tt \mc} we need to determine the domain for
the second argument, while the comparisons we have constrain
only the first one. Thus, we need some technique of {\sl propagating\/}
abstraction domains that we obtained for one subset of integer argument
positions to another subset of integer argument positions. Clearly,
this technique may be seen as a heuristic and it is inapplicable if there
is no interaction between argument positions.

To capture this interaction we draw a graph for each recursive numerical 
predicate, that has the numerical argument positions as vertices and edges 
between vertices that can influence each other. In the case of {\tt \mc} we 
get the graph having an edge between the first argument position and the 
second one.

Let $\pi$ be a permutation of the vertices of a connected component of this 
graph. Define $\pi {\cal D}$ to be the result of replacing each occurrence
of $arg_i$ in ${\cal D}$ by $arg_{\pi (i)}$. 
Consider the Cartesian product of all abstract domains 
$\pi {\cal D}$ thus obtained, discarding unsatisfiable conjunctions.
We will call this Cartesian product the {\em extended domain\/} and denote
it by ${\cal E} {\cal D}$. 
In the case of {\tt \mc} we get as 
${\cal E}{\cal D}$ the set of elements 
{\tt \mc($A$,$B$)}, such that $A$ and $B$ are in
 $\{\mbox{\sl small},\mbox{\sl med},\mbox{\sl big}\}$. 

More generally, when there are arithmetic relations (e.g.\ {\tt Y is X+1}) 
between argument positions,
  ${\cal E}{\cal D}$ can contain
new subdomains that can be inferred from those in ${\cal D}$.

\subsection{Abstract interpretation}
\label{AbstrInterp}
In this section we use the integer abstractions obtained earlier to
classify, in a finite fashion, all possible answers to queries. This analysis
can be skipped in  simple cases (just as in TermiLog 
constraint inference can be skipped when not needed), but is necessary 
in more complicated cases, like {\tt \mc}. Most examples encountered in
practice do not need this analysis.

The basic idea is as follows: define an abstraction domain and perform a
bottom-up constraints inference. 

The abstraction domain that should be defined is a refinement of the 
abstraction domain we defined in Subsection~\ref{IntDomain}. 
There we considered only recursive clauses, since non-recursive clauses do not
affect the query-mapping pairs.
On the other hand, when trying to infer constraints that hold for 
answers of the
program we should consider non-recursive clauses as well. In this way
using one of the techniques presented in the previous subsection both 
for the recursive and the non-recursive clauses 
an abstraction domain $\tilde{\cal D}$ is obtained.
Clearly, $\tilde{\cal D}$ is a refinement of ${\cal D}$.  

\begin{example}
For {\tt \mc} we obtain that the elements of $\tilde{\cal D}$ are the intersections
of the elements in
$ {\cal E}{\cal D}$ (see the end of Subsection~\ref{propagating})with the constraint in the non-recursive clause and its negation.
\end{example}

\begin{example}
\label{Dtilde}
Continuing the {\tt mod}-example we considered in Example~\ref{mod} and 
considering the non-recursive clause for {\tt mod} as well, we
obtain by collecting
comparisons $\tilde{\cal C} = \{arg1\geq arg2, arg2>0, arg1< arg2,
arg3<arg2,arg1 \geq 0,
 arg1\leq arg3, arg1\geq arg3\}$ and, thus, $\tilde{\cal D}$ consists of all 
pairs $(\tt{mod},c)$ for $c$ a satisfiable
element of $2^{\tilde{\cal C}}$.
\eat{
\begin{eqnarray*}
& \tilde{\cal D} = & \{
(\mbox{\tt mod},\mbox{\sl arg1}>\mbox{\sl arg2} \;\&\; \mbox{\sl arg2}>0
 \;\&\; \mbox{\sl arg1} = \mbox{\sl arg3}), \\
&& (\mbox{\tt mod},\mbox{\sl arg1}>\mbox{\sl arg2} \;\&\; \mbox{\sl arg2}>0
 \;\&\; \mbox{\sl arg1} < \mbox{\sl arg3}), \\
&& (\mbox{\tt mod},\mbox{\sl arg1}>\mbox{\sl arg2} \;\&\; \mbox{\sl arg2}>0
 \;\&\; \mbox{\sl arg1} > \mbox{\sl arg3}), \\
&& (\mbox{\tt mod},\mbox{\sl arg1}>\mbox{\sl arg2} \;\&\; \mbox{\sl arg2}\leq 0
 \;\&\; \mbox{\sl arg1} = \mbox{\sl arg3}), \\
&& (\mbox{\tt mod},\mbox{\sl arg1}>\mbox{\sl arg2} \;\&\; \mbox{\sl arg2}\leq 0
 \;\&\; \mbox{\sl arg1} < \mbox{\sl arg3}), \\
&& (\mbox{\tt mod},\mbox{\sl arg1}>\mbox{\sl arg2} \;\&\; \mbox{\sl arg2}\leq 0
 \;\&\; \mbox{\sl arg1} > \mbox{\sl arg3}), \\
&& (\mbox{\tt mod},\mbox{\sl arg1}\leq \mbox{\sl arg2} \;\&\; \mbox{\sl arg2}>0
 \;\&\; \mbox{\sl arg1} = \mbox{\sl arg3}), \\
&& (\mbox{\tt mod},\mbox{\sl arg1}\leq \mbox{\sl arg2} \;\&\; \mbox{\sl arg2}>0
 \;\&\; \mbox{\sl arg1} < \mbox{\sl arg3}), \\
&& (\mbox{\tt mod},\mbox{\sl arg1}\leq \mbox{\sl arg2} \;\&\; \mbox{\sl arg2}>0
 \;\&\; \mbox{\sl arg1} > \mbox{\sl arg3}), \\
&& (\mbox{\tt mod},\mbox{\sl arg1}\leq \mbox{\sl arg2} \;\&\; \mbox{\sl arg2}\leq 0 \;\&\; \mbox{\sl arg1} = \mbox{\sl arg3}), \\
&& (\mbox{\tt mod},\mbox{\sl arg1}\leq \mbox{\sl arg2} \;\&\; \mbox{\sl arg2}\leq 0 \;\&\; \mbox{\sl arg1} < \mbox{\sl arg3}), \\
&& (\mbox{\tt mod},\mbox{\sl arg1}\leq \mbox{\sl arg2} \;\&\; \mbox{\sl arg2}\leq 0 \;\&\; \mbox{\sl arg1} > \mbox{\sl arg3})\}
\end{eqnarray*}
}
\end{example}

Given a program $P$, let $\cal B$ be the corresponding extended Herbrand base,
where we assume that arguments in numerical positions are integers. 
Let $T_{P}$ be the immediate consequence operator. 
Consider the Galois connection provided by the
abstraction function $\alpha: {\cal B} \rightarrow \tilde{\cal D}$ 
and the  concretization function $\gamma: \tilde{\cal D} \rightarrow {\cal B}$ 
defined as follows:
The abstraction $\alpha$ of an element in ${\cal B}$ is the pair from
 $\tilde{\cal D}$
that characterizes it. The concretization $\gamma$ of an element in 
$\tilde{\cal D}$ is the set of all atoms in ${\cal B}$ that satisfy it.
Note that $\alpha$ and $\gamma$ form a Galois connection due to the 
disjointness of the elements of $\tilde{\cal D}$.

Using the Fixpoint Abstraction Theorem (cf.\ \cite{Cousot:Cousot}) we get that
\[\alpha \left ( \bigcup _{n=1}^{\infty}T_{P}^{n}(\emptyset ) \right ) \subseteq \bigcup _{n=1}^{\infty} (\alpha \circ T_{P}\circ \gamma )^{n}(\emptyset)\]
We will take a map ${\tt w}: \tilde{\cal D} \rightarrow\tilde{\cal D}$, that is
a {\em widening\/}~\cite{Cousot:Cousot} of 
$\alpha \circ T_{P}\circ \gamma$ and compute its
fixpoint. Because of the finiteness of $\tilde{\cal D}$ this fixpoint may be
computed bottom-up. 

The abstraction domain 
$\tilde{\cal D}$ describes all possible atoms in the extended 
Herbrand base ${\cal B}$. 
However, it is sufficient for our analysis to describe only
computed answers of the program, i.e., a subset of ${\cal B}$.
Thus, in practice, the computation of the fixpoint can sometimes be simplified as 
follows:
We start with the constraints of the non-recursive clauses. Then we repeatedly
apply the recursive clauses to the set of the constraints obtained thus far, 
but abstract the conclusions to elements of ${\cal D}$. In this way we obtain 
a CLP program that is an abstraction of the original one. This holds in the next example.
The abstraction corresponding to the predicate {\tt p} is denoted
$\tt p_w$.

\begin{example}
Consider once more {\tt \mc}. As claimed above we start from the non-recursive
clause, and obtain that
\begin{eqnarray*}
&& \tt \mc_w(A, B) \imp \{A>100,B=A-10\}.
\end{eqnarray*}
By substituting in the recursive clause of {\tt \mc} we obtain the following
\begin{eqnarray*}
&& \tt \mc(X,Y) \imp X\leq 100,\;Z1\;\;is\;\;X+11, Z1>100,\\
&& \tt \hspace{1.5in} \;Z2\;\;is\;\;Z1-10, Z2>100, \;Y\;\;is\;\;Z2-10.
\end{eqnarray*}
By simple computation we discover that in this case {\tt X} is 100,
and {\tt Y} is 91. However, in order to guarantee the termination of
the inference process we do not infer the precise constraint
$\{X = 100, Y = 91\}$, but its abstraction, i.e., an atom 
$\tt \mc_w({med}, {med})$. Repeatedly applying the
procedure described, we obtain an additional answer 
$\tt \mc_w({small}, {med})$. 

More formally, the following SICStus Prolog CLP(R) program performs the 
bottom-up construction of the abstracted program, as described above.
We use the auxiliary predicate {\tt in/2} to denote a membership in ${\cal D}$
and the auxiliary predicate {\tt e\_in/2} to denote a membership in 
the extended domain ${\cal E}{\cal D}$.
\begin{eqnarray*}
&& \tt \imp use\_module(library(clpr)).\\
&& \tt \imp use\_module(library(terms)).\\
&& \tt \imp dynamic(\mc_w/2).\\
&& \\
&& \tt in(X,big) \imp \{X> 100\}.\\
&& \tt in(X,med) \imp \{X>89,X\leq 100\}.\\
&& \tt in(X,small) \imp \{X\leq 89\}.\\
&& \\
&& \tt e\_in((X,Y),(XX,YY))\imp in(X,XX), in(Y,YY).\\
&& \\
&& \tt \mc_w(X,Y) \imp \{X>100, Y=X-10\}.\\
&& \\
&& \tt assert\_if\_new((H\imp B)) \imp \backslash\!+ (clause(H1,B1),\\
&& \tt\hspace{4cm}unify\_with\_occurs\_check((H,B),(H1,B1))),\\
&& \tt\hspace{4cm} assert((H\imp B)). \\
&& \\
&& \tt deduce \imp \{X\leq 100,Z=X+11\},\mc_w(Z,Z1),\\
&& \tt\hspace{1.2cm} \mc_w(Z1,Y),e\_in((X,Y),(XX,YY)),\\
&& \tt\hspace{1.2cm} assert\_if\_new((\mc_w(A,B) \imp e\_in((A,B),(XX,YY)))),\\
&& \tt\hspace{1.2cm} deduce.\\
&& \tt deduce.
\end{eqnarray*}
The resulting abstracted program is
\begin{eqnarray*}
&& \tt \mc_w(A, B) \imp \{A>100,B=A-10\}.\\
&& \tt \mc_w(A, B) \imp e\_in((A,B), (med, med)).\\
&& \tt \mc_w(A, B) \imp e\_in((A,B), (small, med)).
\end{eqnarray*}
Since we assumed that the query was of the form {\tt \mc($i$,$f$)} we can view
these abstractions as implications of constraints like
$\mbox{\sl arg1}\leq 89$  implies $89<\mbox{\sl arg2}\leq 100$. We also point 
out that the resulting abstracted program coincides with the results obtained
by the theoretic reasoning above.
\end{example}

As an additional example consider 
the computation of the {\sl gcd} according to Euclid's 
algorithm. Proving termination is not trivial, even if the successor notation 
is 
used. In~\cite{Lindenstrauss:Sagiv:Serebrenik:L} only applying a special 
technique allowed to do this.

\begin{example}
Consider the following program and the query {\tt gcd($i$,$i$,$f$)}.
\begin{eqnarray*}
&& \tt gcd(X,0,X) \imp X>0. \\
&& \tt gcd(X,Y,Z) \imp  Y>0,  mod(X,Y,U),  gcd(Y,U,Z).\\
&& \\                          
&& \tt mod(A,B,C) \imp A\geq B, B>0, D\;is\;A-B, mod(D,B,C).\\
&& \tt mod(A,B,C) \imp A< B, A \geq 0, A = C.
\end{eqnarray*}
In this example we have two nested integer loops represented by the predicates 
{\tt mod} and {\tt gcd}. We would like to use the information obtained from 
the abstract interpretation of {\tt mod}
to find the relation between the {\tt gcd}-atoms in the recursive clause.
Thus, during the bottom-up inference 
process we  abstract the conclusions to  elements of
${\tilde{\cal D}_{\tt mod}}$, as it was evaluated in Example~\ref{Dtilde}.
Using this technique  
we get that if $mod(X,Y,Z)$ holds then always $Z<Y$ holds, and this is what is needed to prove the
termination of $gcd$.
\eat{
\begin{eqnarray*}
&& \tt mod_w(X,Y,Z) \imp \{ X\leq Y, X = Z\}.\\
&& \tt mod_w(X,Y,Z) \imp \{ X > Y, Y > 0 \}.
\end{eqnarray*}
That is, two possibilities for the constraints of {\tt mod} are obtained. 
Substituting
these constraints in the rules for {\tt gcd} we can prove termination of the query.
}
\end{example}

\eat{
The approach presented in this subsection is
a novel approach compared to the inter-argument relations that were 
used previously in termination analysis~\cite{Ullman:van:Gelder,Brodsky:Sagiv:2,Plumer:Book,Vershaetse:DeSchreye,Vershaetse:DeSchreye:Deriving:Linear:Size:Relations,Mesnard:Ganascia,DeSchreye:Decorte:NeverEndingStory,Benoy:King,Lindenstrauss:Sagiv,Codish:Taboch}---instead of comparing the sizes of 
arguments the ``if \ldots then \ldots'' expressions are considered, making
the analysis more powerful.
}

\subsection{Query-mapping pairs}
\label{QM}
In this subsection we extend the query-mapping pairs technique 
to programs having numerical arguments.
We  assume that a norm is defined for all arguments.

\eat{
Queries are given as constrained abstract atoms. 
More formally, suppose a partial branch in the LD-tree~\cite{Apt:Book} is given, 
going from node  $\leftarrow A_1,\ldots, A_n $
to node $ \leftarrow B_1,\ldots, B_k$ ,
and let $\theta$ be the composition of the substitutions along 
this branch. Assume also that the atom $B_1$ came into being from the
resolution on $A_1$. In the query-mapping pair corresponding to this
branch, the query is the abstraction of $A_1$ and the
mapping is a quadruple---the domain, the range, arcs and edges.
The domain is the abstraction of $A_1\theta$,
the range is the abstraction of $B_1$ and arcs and edges represent
norm order relations between the nodes of the domain and the range. 
}

We start with the construction of the original query-mapping pairs, but for atoms 
(in the query, domain or range) that are part of integer loops we also add the
appropriate numerical constraints from the integer abstraction domain 
(remember that there
is only a finite number of elements in the integer abstraction domain).

We also
add numerical arcs and edges between numerical argument positions. 
These arcs and edges are added if 
numerical inequalities and equalities between the arguments can be deduced.
Deduction of numerical edges and arcs is usually done by considering the 
clauses. However, if a subquery $q$ unifies with a head of a clause of the 
form $A\imp B_1,\ldots,B_k,\ldots,B_n$ and we want to know the relation 
between $q$ and $B_k$ (under appropriate substitutions), we {\em may} use the
results of the abstract interpretation to conclude numerical constraints for
$B_1,\ldots,B_{k-1}$. The reason is that if we arrive at $B_k$, this
means that we have proved $B_1,\ldots,B_{k-1}$ 
(under appropriate substitutions). All query-mapping pairs deduced in this 
way are then repeatedly composed. The process terminates because there is
a finite number of query-mapping pairs. 

A query-mapping pair is called {\em circular} if the
query coincides with the range. The initial query terminates if for 
every circular query-mapping pair one of the following conditions
holds:
\begin{itemize}
\item The circular pair meets 
the requirements of the termination test of Theorem~\ref{qm-th}.
\item 
There is a non-negative termination function for which we can prove a decrease
from the domain to the range using the numerical edges and arcs and the 
constraints of the domain and range.
\end{itemize}

Two questions remain: how does one automate the guessing of the function, and how does one prove that it decreases. 
Our heuristic for guessing a termination function is based on the inequalities 
appearing in the abstract constraints.  Each inequality of the form 
{\sl Exp1}$\;\rho\;${\sl Exp2} where $\rho$ is one of $\{>,\geq\}$ suggests 
a function $\mbox{\sl Exp1} - \mbox{\sl Exp2}$. 

The common approach to termination analysis is to find 
{\em one} termination function that decreases over all possible execution paths. This leads to complicated termination functions. Our approach allows one
to guess a number of relatively simple termination functions, each suited 
to its query-mapping pair. When termination functions are simple to find, the 
guessing process can be performed automatically.

After the termination function is guessed, its decrease must be proved.
Let $V_1,\ldots,V_n$ denote numerical argument positions in the domain and
$U_1,\ldots,U_n$ the corresponding numerical argument positions in the range
of the query-mapping pair. First, edges of the query-mapping pair 
are translated to equalities and arcs, to inequalities between these variables.
Second, the atom constraints for the $V$'s
and for the $U$'s are added.
Third, let $\varphi$ be a termination function. We would like to check that
$\varphi(V_1,\ldots,V_n) > \varphi(U_1,\ldots,U_n)$ is implied by the constraints. Thus, we add the negation of this claim to the collection of the constraints
and check for unsatisfiability. Since termination functions are linear,
CLP-techniques,
such as CLP(R)~\cite{Jaffar:Maher} and CLP(Q,R)~\cite{CLP:Manual},
are robust enough to obtain the desired contradiction. Note however, 
that if more powerful constraints solvers are used, non-linear termination
functions may be considered.

To be more concrete:
\begin{example}
\label{difficult:loops}
Consider the following program with query {\tt p($i$,$i$)}.
\begin{eqnarray*}
&& \tt p(0,\_).\\
&& \tt p(X,Y)\imp X>0,X<Y,X1\;\;is\;\;X+1, p(X1,Y).\\
&& \tt p(X,Y)\imp X>0,X\geq Y,X1\;\;is\;\;X-5, Y1\;\;is\;\;Y-1, p(X1,Y1).
\end{eqnarray*}

We get, among others, the circular query-mapping pair having the query 
({\tt p($i$,$i$)},$\{\mbox{\sl arg1}>0,\mbox{\sl arg1}<\mbox{\sl arg2}\}$) 
and the mapping given in Figure~\ref{qmdl}. The termination function 
derived for the circular 
query-mapping pair is {\sl arg2}$-${\sl arg1}. In this case, 
we get from the arc and the edge 
the constraints: $V_1<U_1, V_2=U_2$. 
We also have that $V_1>0, U_1>0, V_1<V_2, U_1<U_2$. We would like to prove that
$V_2-V_1>U_2-U_1$ is implied. Thus, we add $V_2-V_1\leq U_2-U_1$ to the set of
constraints and CLP-tools
easily prove unsatisfiability, and thus, that the required implication holds.
\end{example}
\wgefig{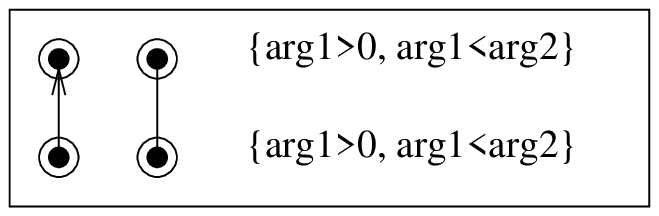}{Mapping for {\tt p}}{qmdl}{0.40\textwidth}

In the case of the 91-function the mappings are given in Figure~\ref{qm91}. 
(We omit the queries from the query-mapping pairs, since
they are identical to the corresponding domains.)
\wgefig{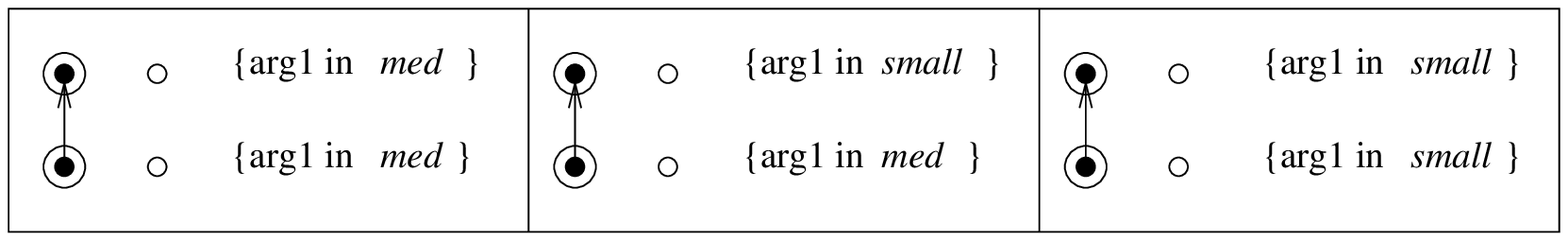}{Mappings for McCarthy's 91 function}{qm91}{0.95\textwidth}

In the examples above there were no interargument relations of the type considered in~\cite{Lindenstrauss:Sagiv}.
However, this need not 
be the case in general. 
\begin{example}
\label{both:types:of:pairs}
Consider the following program with the query 
{\tt q($b$,$b$,$i$)} and the term-size norm. 
\begin{eqnarray*}
&(1)&\tt q(s(X),X,\_).\\
&(2)&\tt q(s(X),X,N)\imp N>0, N1\;\;is\;\;N-1, q(s(X),X,N1).\\
&(3)&\tt q(s(s(X)),Z,N)\imp N=<0, N1\;\;is\;\;N-1, q(s(X),Y,N1),q(Y,Z,N1).
\end{eqnarray*}
Note that constraint inference is an essential step for
proving termination---in order to infer that there is a norm decrease in the first argument
between the head of (3) and the second recursive call (i.e. $\parallel 
s(s(X))\parallel >
\parallel Y\parallel$), one should infer that the second argument in {\tt q} is less than the
first with respect to the norm (i.e. $\parallel s(X)\parallel
>\parallel Y\parallel$). We get
among others circular query-mapping pairs having the mappings presented 
in Figure~\ref{qmq}. The queries of the mappings coincide with the 
corresponding domains. In the first mapping termination follows from the 
decrease in the third
argument and the termination function {\sl arg3}$>0$. 
In the second mapping termination
follows from the norm decrease in the first argument.
\wgefig{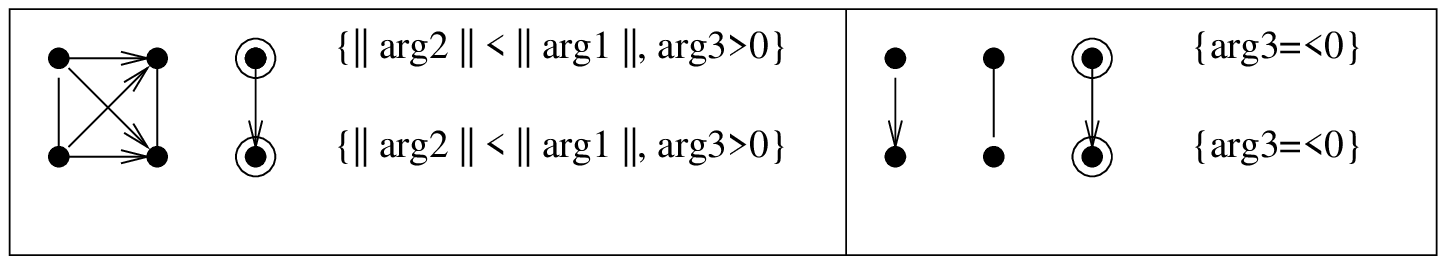}{Mappings for {\tt q}}{qmq}{0.80\textwidth}
\end{example}

\subsection{The Extended Algorithm}
\label{Algorithm}
In this section we combine all the techniques suggested so far. The complete algorithm {\sf Analyze\_Termination} is presented in
Figure~\ref{algo}. Each step corresponds to one
of the previous sections. 

\begin{figure}[htb]
\begin{center}
\fbox{
\parbox{6in}{
\begin{tabbing}
AAAA \= AAAAA\= AAAAA\= AAAAA\= AAAAA\= AAAAA\= AAAAA\=\kill
{\bf Algorithm} \>\> {\sf Analyze\_Termination}\\
{\bf Input}     \>\> A query pattern $q$ and a Prolog program $P$\\
{\bf Output}    \>\> {\sf YES}, if termination is guaranteed\\
		\>\> {\sf NO}, if no termination proof was found\\
\\
\ \,(1) \> {\bf Guess} and {\bf verify} numerical argument positions;\\
\ \,(2) \> {\bf Compute} the integer abstraction domain;\\
\ \,(3) \> {\bf Compute} abstractions of answers to queries (optional);\\
\ \,(4) \> {\bf Compute} ordinary and numerical query-mapping pairs;\\
\ \,(5) \> {\bf For each} circular query-mapping pair {\bf do}:\\
\ \,(6) \> 	\> {\bf If} its circular variant has a forward positive cycle {\bf then}\\
\ \,(7) \>      \>      \> {\bf Continue};\\
\ \,(8) \>      \>{\bf If} the query-mapping pair is numerical {\bf then}\\
\ \,(9) \>     	\>      \> {\bf Guess} bounded termination function;\\
(10)    \>	\>	\> {\bf Traverse} the query-mapping pair {\bf and  compute} values\\
     \>		\>	\>	\>of the termination function;\\
(11) \> 	\> 	\> {\bf If} the termination function decreases monotonically {\bf then}\\ 
(12) \> 	\>      \> 	\> {\bf Continue};\\
(13) \> 	\> {\bf Return} {\sf NO}; \\
(14) \> {\bf Return} {\sf YES}.
\end{tabbing}
}}
\end{center}
\caption{Termination Analysis Algorithm}
\label{algo}
\end{figure}

Note that Step 3, computing the abstractions of answers to queries, is 
optional. If the algorithm returns {\sf NO}
it may be re-run either with Step 3 included or with a different integer 
abstraction domain. 

The {\sf Analyze\_Termination} algorithm is sound:
\begin{theorem}
Let $P$ be a program and $q$ a query pattern.
\begin{itemize}
\item {\sf Analyze\_Termination($P$, $q$)} terminates.
\item If {\sf Analyze\_Termination($P$, $q$)} reports {\sf YES} then,
for every query $Q$ matching the pattern $q$, the LD-tree of $Q$ w.r.t.\ $P$ is finite.
\end{itemize}
\end{theorem}

Work is being done now to implement the ideas in this section, and thus to be 
able to deal with programs for which termination depends on the
behavior of arithmetic predicates.
\section{Conclusion and Generalizations}
We have seen the usefulness of the query mapping-pairs approach for proving
termination of queries to logic programs by using symbolic linear norm 
relations between arguments and also by comparing numerical arguments.

In the query-mapping pairs method as outlined above there are two crucial
elements:
\begin{enumerate}
\item There is a finite number of abstractions of atoms in
subgoals of the LD-tree. This ensures that $\cal A$ is finite. 
\item Arcs represent an order.
\end{enumerate}
This suggests two directions for generalization---using different abstractions
and using different orders.

\subsection{Using Different Abstractions of Terms and a Linear Norm}
We can use the original 
query-mapping pairs as before with the only difference that we'll
abstract nodes not to just black and white ones but to a larger, though
finite, set. For instance if we have a program
\begin{eqnarray*}
&& \tt p(1) \imp \{ \mbox{\tt infinite loop}\}.\\
&& \tt p(0).
\end{eqnarray*}
\noindent and take the term-size norm
and a query $p(bound)$, the query-mapping algorithm will say that there
may be non-termination. However, we can use the abstractions $1,g,f$, where $g$
means any ground term that is not $1$ and $f$ means any  term,
and apply the above algorithm, with the only difference being in the
unification of the abstractions.
 In the  algorithm in Subsection~\ref{qm-pairs-algo} we used unification of
abstractions in two places---when adjusting the weighted rule graph to the
instantiation pattern of the query and when composing query-mapping pairs.
In those cases the result of the unification of two nodes of which at
least one was black resulted in a black node, and the unification of two
white nodes resulted in a white node. In the present case
 $g$ and $1$ will not unify so we will be able to prove that
  a query $p(g)$ terminates.

  \begin{observation}
  The original query-mapping pairs algorithm remains valid if we abstract arguments
  of atoms in the LD-tree to elements of any finite set of abstractions,
 as long as we include a sound procedure for unification of these
  abstractions.
  \end{observation}

\subsection{Using Norms that Involve Ordinal Numbers}
There are programs for which the use of linear norms is not sufficient.
The following program performs repeated differentiation.
\begin{eqnarray*}
&& \tt d(deriv(t),1).\\
&& \tt d(deriv(A),0) \imp number(A).\\
&& \tt d(deriv(X+Y),L+M)  \imp  d(deriv(X),L),d(deriv(Y),M).\\
&& \tt d(deriv(X*Y),(X*L+Y*M))  \imp  d(deriv(X),M),d(deriv(Y),L).\\
&& \tt d(deriv(deriv(X)),L)  \imp  d(deriv(X),M),d(deriv(M),L).
\end{eqnarray*}

In this case one can show that for no choice of constants in the definition
of the linear norm will it be possible to prove termination of
$d(ground,free)$.
However, we can use the query-mapping pairs method with the abstraction of
arguments to ground and non-ground,
but use a norm that associates with each term an ordinal number in the
following way ($\omega$ denotes, of course, the first infinite ordinal):
\[\|deriv(X)\| = \omega + \| X\|\]
\[\|X+Y\| = \|X\| \oplus \|Y\| + 2\]
\[\|X*Y\| = \|X\| \oplus \|Y\| + 2\]
where $(n_1 \omega + k_1) \oplus (n_2 \omega + k_2)$ for non-negative integers
$n_1,n_2,k_1,k_2$ is defined as 
$\mbox{\sl max}(n_1, n_2)\omega + (k_1 + k_2)$ and + is a usual addition of ordinal numbers. 
\begin{observation}
The original query-mapping algorithm remains valid if we replace,
in the computation of
norms, integers by ordinal numbers with the operations defined above.
\end{observation}

\noindent
{\bf Acknowledgement:} We are very grateful to the anonymous referees for their careful
reading and helpful suggestions.

\bibliography{main}
\bibliographystyle{plain}

\end{document}